\documentclass[%
 aip,
 amsmath,amssymb,
preprint,%
]{revtex4-1}

\usepackage{graphicx}
\usepackage{dcolumn}
\usepackage{bm}

\usepackage[utf8]{inputenc}
\usepackage[T1]{fontenc}
\usepackage{mathptmx}
\usepackage{etoolbox}
\usepackage{comment}

\usepackage{xcolor}
\usepackage{soul}
\definecolor{customgreen}{HTML}{009933}

\usepackage{tabularx}
\usepackage{multirow}
\usepackage{verbatim}

\definecolor{darkpink}{rgb}{0.8, 0.2, 0.3}

\definecolor{olivegreen}{rgb}{0.33, 0.42, 0.18}

\begin{document}

\title{Collisional-radiative data for tokamak disruption mitigation modeling}

\author{Prashant Sharma}
\email{phyprashant@gmail.com}
\affiliation{Theoretical Division, Los Alamos National Laboratory, Los Alamos, NM 87545}

\author{Christopher J. Fontes}
\email{cjf@lanl.gov}
\affiliation{Computational Physics Division, Los Alamos National Laboratory, Los Alamos, NM 87545}

\author{Dmitry V. Fursa}
\affiliation{Department of Physics and Astronomy, Curtin University, Perth, Western Australia 6102, Australia}

\author{Igor Bray}
\affiliation{Department of Physics and Astronomy, Curtin University, Perth, Western Australia 6102, Australia}

\author{Mark Zammit}
\affiliation{Theoretical Division, Los Alamos National Laboratory, Los Alamos, NM 87545}

\author{James Colgan}
\affiliation{Theoretical Division, Los Alamos National Laboratory, Los Alamos, NM 87545}

\author{Hyun-Kyung Chung}
\affiliation{Korea Institute of Fusion Energy,  169-148 Gwahak-ro, Yuseong-gu, Daejeon 34133, Korea}

\author{Nathan Garland}
\affiliation{Theoretical Division, Los Alamos National Laboratory, Los Alamos, NM 87545}
\affiliation{Queensland Quantum and Advanced Technologies Research Institute, Griffith University, Nathan, Queensland 4111, Australia}
\affiliation{School of Environment and Science, Griffith University, Nathan, Queensland 4111, Australia}

\author{Xian-Zhu Tang}
\email{xtang@lanl.gov}
\affiliation{Theoretical Division, Los Alamos National Laboratory, Los Alamos, NM 87545}


\begin{abstract}
Effective tokamak disruption mitigation is crucial for ensuring the
safety and integrity of fusion power reactors.  Accurate
collisional-radiative (CR) modeling of a radiative plasma is a
critical component in predictive disruption mitigation design.  In
this paper, we focus on quasi-steady-state CR modeling applicable to
the current quench phase of a tokamak disruption.  We employ the
ATOMIC collisional-radiative code from the Los Alamos suite and the newly
developed Fusion Collisional-Radiative (FCR) code to model the atomic
processes, providing high-fidelity data for radiative power loss, as
well as average and effective charge states for hydrogen, helium,
neon, and argon plasma species over a wide range of tokamak-relevant
electron temperatures and electron densities. Fine-structure-resolved
CR models are used for hydrogen and helium plasma species, while
configuration-average CR models are implemented for neon and argon
plasma species. The calculated values are compared with the
superconfiguration CR model (FLYCHK) and the commonly used coronal
equilibrium approximation to demonstrate the advantages and
limitations of each model. To facilitate coupling of high-fidelity CR data to plasma simulation models, we represent the ATOMIC/FCR results over the relevant plasma parameter range using a smooth tensor product B-spline surface in electron temperature and electron density. This approach yields compact coefficient tables that can be evaluated efficiently while preserving spline smoothness across the domain. These data were previously used to examine ways to minimize runaway electrons in a tokamak current quench, and they are now made available in easy-to-use forms for community use and benchmarking.
\end{abstract}

\maketitle


\section{Introduction}
A tokamak disruption begins with a rapid thermal quench (TQ) in which
the plasma thermal energy is mostly lost on the order of a millisecond
(ms) or so~\cite{Hender:2007,riccardo2005timescale}, and proceeds to a
much slower current quench (CQ) phase in which the plasma current is
dissipated over a timescale on the order of 100 ms or so, as
anticipated for ITER~\cite{Hender:2007, Hollmann-etal-PoP-2015}. For a
typical short TQ duration, a greatly increased thermal load at the
divertor and neighboring first wall is expected, compared with
steady-state reactor operation. To mitigate this, a proposed tactic is
to inject impurities with relatively high atomic numbers, such as argon
and neon, into the disrupting plasma to promote radiative energy loss
that can disperse the plasma thermal load over the entire chamber
wall~\cite{Whyte:2002,lehnen2015disruptions}. These high-$Z$
impurities also play a critical role in the post-TQ plasma, where
Ohmic-to-runaway current conversion can take place to form a runaway
current plateau \cite{vallhagen2020runaway}. As the plasma energy
losses via parallel and perpendicular transport (with respect to the
toroidally dominant magnetic field) become ineffective in a
low-temperature
plasma~\cite{Li-Zhang-Tang-TQ-2022,ward1992impurity,breizman2019physics},
the post-TQ plasma power balance is mostly set by the competition
between radiative cooling and the collisional heating associated with
the plasma current dissipation~\cite{mcdevitt2022}. Accurate modeling
of the underlying atomic processes in a disrupting tokamak plasma,
describing the charged state populations of all species and radiative
cooling rates for electrons, is thus essential for disruption
mitigation design \cite{mcdevitt2022}. Depending on the electron
temperature ($T_e$) and density ($n_e$) in the plasma, one can use
different levels of approximation in defining plasma models to deduce
the plasma parameters. For example, plasmas with high temperatures and
low densities can be modeled using the collisional-ionization
equilibrium or coronal-equilibrium (CE) model, whereas for
intermediate temperatures and densities, a general
collisional-radiative (CR) model can be used. A detailed discussion of
these models is presented in the following sections. Both models
require the solution of a set of equations to determine the ionization
and excitation states of the plasma, but they vary in the kinds of
atomic processes they include and, consequently, in the accuracy and
amount of computation required.

For tokamak disruption mitigation modeling, this balance between
physics fidelity and computational cost informs practical
compromises. As an example, if we consider well-studied JET scenarios
where the TQ phase can last from 0.05 to 1~ms
\cite{hollmann2005measurements,marin2022integrated}, an injected
high-$Z$ impurity neon (high compared to hydrogen and helium, but
still much lower compared to tungsten from the divertor plate on
ITER) would take up to 1~ms to reach collisional-radiative
equilibrium even at fixed $T_e$. The rapid cooling of the plasma
during a thermal quench thus demands the approach of a time-dependent
CR model, as opposed to a steady-state CR model. The current quench
phase provides a sharp contrast in that the plasma temperature already
settles to a low value where relatively slight variation is possible,
and the current quench duration is around 100~ms, two orders of
magnitude longer than the typical equilibration time in the CR
model. This behavior suggests that a steady-state CR model can be
appropriate here to quasistatically couple the CR physics to
models of the background plasma and possible runaway electron formation. There is a further
refinement in the detailed choice of the steady-state CR model
depending on whether Ohmic-to-runaway conversion can produce strong
runaway electron current. If the runaway conversion is efficient and
runaway current density becomes sufficiently high, the relativistic
correction of cross-sections for collisional excitation and ionization
must be appropriately accounted for
\cite{garland2020impact,garland2022understanding}. In the runaway
avoidance studies, the runaway current is to be minimized
\cite{mcdevitt2022} so one can retain the CR model without the
aforementioned relativistic corrections.

In this paper, we focus on the steady-state CR modeling that is needed
to quasistatically couple the CR physics to plasma modeling during the
current quench phase in which the runaway current density is not high
enough for the relativistic, or quantum electrodynamic (QED),
corrections to be important. This is the operational regime of runaway
avoidance/minimization. As will be shown in
Sec.~\ref{sec:plasma-coupling}, the coupling of the CR model to the
plasma dynamics in such scenarios is straightforward, such that one can
precompute the relevant CR quantities for a prescribed grid of
conditions and then perform the plasma-model coupling via spline-based
evaluation of these CR data. This approach opens
up the possibility of deploying relatively high-fidelity CR models. In
the present work, we use two CR codes, namely ATOMIC \cite{Fontes2015}
and FCR \cite{sharma2026hybrid}. 
While ATOMIC relies on atomic data that are generated with the Los Alamos suite of codes, FCR can make use of additional sources of atomic data, such as convergent
close-coupling (CCC) calculations and community databases. For comparison, we contrast results obtained from these
fine-structure (ATOMIC) and configuration-average (FCR) CR models
 with the corresponding quantities obtained from a widely used superconfiguration CR model (FLYCHK
\cite{CHUNG20053}), as well as from the commonly used coronal equilibrium
model.

The rest of the paper is organized as follows. In
Sec.~\ref{sec:atomic_model}, we present a detailed discussion on the
CR model used in the work. This is followed by
Sec.~\ref{sec:plasma-coupling}, where we outline the motivation and
methodology of applying the CR models to current quench relevant
plasmas. We then discuss the calculated results, i.e., radiative power
loss ($P_{\text{rad}}$), average charge state ($Z_{\text{avg}}$), and
effective charge state ($Z_{\text{eff}}$), of hydrogen, helium, neon,
and argon in Sec.~\ref{sec:atomic_calculations}.  The following
section (Sec.~\ref{sec:model_compare}) contains comparisons with
different models and a discussion of the shortcomings of those
models. Furthermore, a discussion of the B-spline surface fitting methodology is presented in the subsequent section (Sec.~\ref{sec:cubic}) before
concluding with a summary of the present work.

\section{Collisional-Radiative Modeling Approach \label{sec:atomic_model}}
The atomic processes in a tokamak plasma are most comprehensively
described by a CR model, which tracks the population densities of all
charge states, as well as the various ground and excited states in
each charge state. The standard form of a CR model is given
by\cite{fontes2016modern},
\begin{align}
\frac{d \mathbf{n}}{d t} = \mathbf{\mathsf{R}} \cdot \mathbf{n}, 
\end{align}
where $\mathbf{n}$ is the state vector of ground- and excited-state
populations of each ion stage being considered for the target atom(s),
and $\mathbf{\mathsf{R}}$ is the CR rate matrix where each element,
$R_{ij}$, is the sum of transition rates between state $i$ and state
$j$ due to all considered collisional and radiative processes in the
plasma. Common examples include electron impact (de-)excitation and
ionization (recombination) or spontaneous radiative decay. In certain
situations, such as a partially ionized plasma, the process of
charge-exchange between ions and neutrals can be significant enough
that it must be accounted for in the CR rate matrix
\cite{hui2009ion,ali2010critical,liu2022total}. For $R_{ij}$ that have
contributions from collisional processes, the electron distribution
function, $f_e$, enters directly. In cases where $f_e$ can be
approximated with a Maxwellian distribution, the $R_{ij}$ will have an
explicit dependence on electron temperature, $T_e$, and density,
$n_e$. For three-body processes, such as three-body recombination, the
contribution to $R_{ij}$ will have a $n_e^2$ dependence. If
charge-exchange processes are also taken into account, the $R_{ij}$
can have an explicit dependence on $\mathbf{n}.$ When quasi-neutrality
is enforced in a CR calculation at fixed $T_e,$ for example, the
$R_{ij}$ contributions associated with electron-ion collisional
processes would have an implicit dependence on $\mathbf{n}$ through
$n_e$. A more comprehensive discussion on CR models can be found
elsewhere \cite{ralchenko2016modern}.

The computational complexity of CR models is determined by the level of refinement \cite{fontes2016modern} chosen for the atomic energy states of an atom or ion. 
A convenient theoretical concept when discussing atomic energy levels is the notion of a configuration. For the purposes of this work, we denote a configuration by a collection of orbitals (alternatively referred to as subshells), denoted by $nl$, where $n$ is the principal quantum number and $l$ is the orbital angular momentum quantum number. In addition, each subshell contains a particular number of electrons, $w$, which is referred to as the occupation number. Thus, a complete description of a particular subshell is denoted by $nl^w$, and a particular configuration that resides within a given ion stage is specified by a collection of $nl^w$ values.
The most sophisticated level of refinement at which we can calculate
the energies of the atomic states is referred to as the fine-structure approach. In this case, the angular momenta of the orbital electrons are coupled to produce a total angular momentum $J$ for each atomic energy level. A more approximate method for calculating energies is to consider the configurations themselves as the atomic states of interest. This approach is referred to as the  configuration-average approximation. Further averaging over the $l$ quantum number leads to the so-called superconfiguration approximation, which employs an $n^w$, rather than $nl^w$, type of notation to denote the super-shells that make up a particular superconfiguration.
As the atomic number increases, the size of the state vector $\mathbf{n}$ for these different levels of refinement can range from 10 to 10$^9$, or even higher. Constructing reliable and computationally tractable CR models for diverse applications requires understanding the consequences of each departure from the ideal state space, informed by the intended use of the model. For instance, a model designed for inline use in radiative hydrodynamics codes demands speed and validity over a wide range of conditions, whereas a model intended for high-resolution spectroscopic data analysis requires highly accurate state structure and populations. Thus,  the requirement of balancing state-space completeness, state detail, and computational tractability significantly influences the choice of the CR model, necessitating a trade-off between the fidelity of the physical representation and the computational cost. 

\section{Coupling the CR data to plasma modeling\label{sec:plasma-coupling}}
There are relatively few atomic elements involved in the current
quench of a tokamak disruption. A steady-state CR model calculation of
the plasma parameters, such as $Z_{\text{avg}}$, $Z_{\text{eff}}$, and
$P_{\text{rad}}$, for a mix of atomic species $\alpha$, can provide a
baseline for atomic processes in the fusion plasma, especially in
regions away from the boundary where particle recycling enters the CR
model directly, and when the temporal plasma dynamics is slow compared
with the CR equilibration time, which is likely satisfied in the
current quench phase of a tokamak disruption. These plasma parameters
hold significant importance in comprehending plasma dynamics. For
instance, accurate knowledge of radiative power loss due to impurities
is crucial for managing power dissipation and achieving ignition
conditions, given that a magnetic fusion plasma contains various
impurities. Concurrently, estimates of the average charge state and
effective charge state are critical in calculating the electron
density assuming the quasi-neutrality condition and the electron-ion
and electron-atom conductivities \cite{hirshman1978neoclassical},
respectively. If ion-ion
collisional processes are ignored in the CR model, one can construct
$(Z_{\text{avg}}, Z_{\text{eff}}, P_{\text{rad}})$ from single atomic
species CR calculation of $(Z_{\text{avg}}^\alpha,
Z_{\text{eff}}^\alpha, p_{\text{rad}}^\alpha)$.  For an atomic mix of
$\{n_\alpha\},$ each atomic species can have different charge state
population distribution $\{ n_\alpha^Z \}$, where $Z$ is the total
charge of the atom/ion species. The average charge state for species
$\alpha$ is defined as,
\begin{align}
Z_{\text{avg}}^\alpha \equiv \frac{\sum_{Z=0}^{Z_\alpha} Z n_\alpha^Z}{n_\alpha}, 
\end{align}
with the atomic density of species $\alpha$,
\begin{align}
n_\alpha = \sum_{Z=0}^{Z_\alpha} n_\alpha^Z. 
\end{align}
Summing over all the ion species, the average charge state of the mixed plasma is,
\begin{align}
Z_{\text{avg}} \equiv \frac{\sum_{\alpha} Z_{\text{avg}}^\alpha n_\alpha}{n_i},
\end{align}
where the total atomic density,
\begin{align}
n_i = \sum_\alpha n_\alpha.
\end{align}
The effective charge state for species $\alpha$ is defined as,
\begin{align}
Z_{\text{eff}}^\alpha \equiv \frac{\sum_{Z=0}^{Z_\alpha} Z^2 n_\alpha^Z}{Z_{\text{avg}}^\alpha n_\alpha}.
\label{eq:Zeff-species}
\end{align}
For the entire ion mix, the effective charge state is defined as,
\begin{align}
  Z_{\text{eff}} \equiv \frac{\sum_\alpha \sum_{Z=0}^{Z_\alpha} Z^2 n_\alpha^Z}{\sum_\alpha\sum_{Z=0}^{Z_\alpha} Z n_\alpha^Z}.
\end{align}
This can be computed as,
\begin{align}
  Z_{\text{eff}} = \frac{\sum_\alpha Z_{\text{eff}}^\alpha Z_{\text{avg}}^\alpha n_\alpha}{\sum_\alpha Z_{\text{avg}}^\alpha n_\alpha}.
\end{align}

It is important to note that in an atomic mix $\{n_\alpha\},$ the free
electron density is given by the constraint of quasi-neutrality,
\begin{align}
n_e = \sum_\alpha \sum_{Z=0}^{Z_\alpha} Z n_\alpha^Z = \sum_\alpha Z_{\text{avg}}^\alpha n_\alpha.
\end{align}
The CR model can predict $Z_{\text{avg}}^\alpha\left(n_e,T_e\right)$
and $Z_{\text{eff}}^\alpha\left(n_e, T_e\right)$ as functions of free
electron density $n_e$ and temperature $T_e.$ This is usually done in
a single ion species CR calculation with imposed $n_e$ and $T_e.$ The
underlying logic is that the ion charge and energy state population,
which we label as $n_\alpha^j$ where $j$ runs through all energy
levels for all charge states of a single atomic species $\alpha$, is
solved in a CR model of the form,
\begin{align}
\frac{d n_\alpha^j}{d t} = \sum_k R_\alpha^{jk} n_\alpha^k, 
\label{eq:CR-model}
\end{align}
where $R_\alpha^{jk}(n_e, T_e)$, the rate matrix, which is a function
of free electron density and temperature, or more precisely, the free
electron distribution function. The rate of binary collisions between
electrons and ions scales with $n_e,$ but three-body collisions, such
as the three-body recombination process, have a rate that scales with
$n_e^2$. This clearly implies that the fractional density
$n_\alpha^Z/n_\alpha$ is mostly a function of $T_e,$ with weak $n_e$
dependence due to the three-body recombination effect, which is more
pronounced in the high $n_e$ limit. As a result,
$Z_{\text{avg}}^\alpha$ and $Z_{\text{eff}}^\alpha$ follow a similarly
weak $n_e$ dependence. From the solution of Eq.~(\ref{eq:CR-model}),
one can compute the radiative power loss rate for atomic species
$\alpha,$ which we denote as $P_{\text{rad}}^\alpha.$ For binary
collisions, $P_{\text{rad}}^\alpha$ would scale linearly with $n_e
n_\alpha,$ and is a strongly varying function of $T_e$. The normalized
radiative power loss rate,
\begin{align}
p_{\text{rad}}^\alpha \equiv \frac{P_{\text{rad}}^\alpha}{n_e n_\alpha}, \label{eq:normalized-prad}
\end{align}
is a strongly varying function of $T_e$ but only has a weak dependence
on $n_e$, except when the three-body recombination becomes important
in the high $n_e$ regime.

In the present work, the way we are going to use the CR steady state prediction of $Z_{\text{avg}}^\alpha(n_e, T_e), Z_{\text{eff}}^\alpha(n_e, T_e),$ and $p_{\text{rad}}^\alpha(n_e, T_e)$ is as follows. For an atomic mix of $\{n_\alpha\}$ at $T_e$, we actually do not know what $n_e$ would be {\em a priori.} So we will solve $n_e$ and $Z_{\text{avg}}$ from the coupled equations at given $T_e,$
\begin{align}
  n_e & = Z_{\text{avg}} \sum_\alpha n_\alpha, \\
  Z_{\text{avg}} \sum_\alpha n_\alpha & = \sum_\alpha Z_{\text{avg}}^\alpha(n_e, T_e) n_\alpha.
\end{align}
Then $n_e$ is substituted into
\begin{align}
  Z_{\text{eff}} & = \frac{\sum_\alpha Z_{\text{eff}}^\alpha(n_e, T_e) Z_{\text{avg}}^\alpha(n_e, T_e) n_\alpha}{\sum_\alpha Z_{\text{avg}}^\alpha(n_e, T_e) n_\alpha}, \\
  P_{\text{rad}} & = \sum_\alpha p_{\text{rad}}^\alpha(n_e, T_e) n_e n_\alpha,
\end{align}
to compute the $Z_{\text{eff}}$ and the total radiative power loss rate $P_{\text{rad}}.$

In summary, we will need to use a CR code, in the case of single atomic species $\alpha,$ to evaluate $Z_{\text{avg}}^\alpha(n_e, T_e),$ $Z_{\text{eff}}^\alpha(n_e, T_e),$ and
$p_{\text{rad}}^\alpha(n_e, T_e)$ for a range of $n_e$ and $T_e.$ For the application of tokamak disruption mitigation, the range of interest is
\begin{align}
  n_e \in \left[ 10^{12}, 10^{17}\right]\, {\rm cm}^{-3},\,\,\,
  T_e \in \left[1, 4\times 10^4\right]\, {\rm eV}. 
\end{align}
It is noteworthy that the atomic species of fusion interest for disruption mitigation are deuterium, tritium, helium, neon, and argon. With this motivation, we have presented the new data for hydrogen, helium, neon, and argon in this study.

\section{ATOMIC data: Hydrogen, Helium, Neon, and Argon \label{sec:atomic_calculations}}

In the present work, the radiative power loss, average charge state, and effective charge state were calculated for four plasma species, i.e., hydrogen, helium, neon, and argon. The results presented in this work were calculated using two codes developed at Los Alamos National Laboratory (LANL): the ATOMIC CR code \cite{Fontes2015} and the Fusion Collisional-Radiative (FCR) code \cite{sharma2026hybrid}. The hydrogen plasma quantities were calculated with the FCR code, which incorporated a combination of two sets of atomic data. The electron-impact excitation and  electron-impact ionization cross-sections were generated with the CCC approach, while the atomic structure data, including energy levels and oscillator strengths, and photoionization cross-sections were generated with the LANL suite of atomic physics codes \cite{Fontes2015}.

For the calculation of helium, neon, and argon plasma quantities, we used the ATOMIC CR code, which relies exclusively on the LANL suite of atomic physics codes for all data calculations \cite{Fontes2015}. For completeness, we provide here a brief description of the codes in the LANL suite. The CATS atomic structure code\cite{cats1988}, based on Cowan's atomic structure codes\cite{cowan1981theory}, was used to generate the necessary wavefunctions, energy levels, and radiative decay rates with the semi-relativistic Hartree-Fock option. The multipurpose ionization code GIPPER\cite{abdallah1994kinetics,clark1991integral} was employed to calculate the electron-impact ionization, photoionization cross-sections, and autoionization rates. The rates of inverse processes, i.e., three-body recombination, radiative recombination, and electron capture, were determined using the principle of detailed balance. The electron-impact excitation data for transitions from low-lying states are calculated using the ACE code\cite{ace1988}, which is based on the distorted-wave method. Electron-impact excitation data for transitions between all other states were calculated using the more approximate plane-wave Born method.

The fine-structure mode was employed in the case of hydrogen and helium, with a maximum principal quantum number of $n=10$. In both cases, singly ionized states are considered up to $10l$, whereas for helium, doubly excited states are included up to $2l\, 10l'$ in the neutral stage. This results in a total of 101 and 1046 fine-structure levels for hydrogen and helium, respectively.

The more approximate configuration-average model was used for the neon and argon cases, with a maximum principal quantum number of $n=8$ and $n=10$, respectively, in order to make the CR calculations more computationally tractable for these complex systems. The neon model was constructed using single-electron permutations from the valence shell of the ground configuration of a given ion stage up to the $n=8$ shell. For ion stages with at least two electrons in the $n=2$ shell of the ground configuration, i.e., Ne$^{0+}$ -- Ne$^{6+}$, double permutations were considered from the $n=2$ shell up to excited configurations of the form $\ldots 3l\,nl'$, where $3 \le n \le 5$ in all cases except for Ne$^{6+}$, for which an extended range of $3 \le n \le 8$ was considered. The same type of double promotions, over the same ion stages, were also considered to create configurations of the form $\ldots 4l\,4l'$. The promotion of $1s$ electrons was also considered for the Ne$^{8+}$ and Ne$^{9+}$ ion stages. The above permutations result in a neon model that contains 1,508 configurations.

The argon model was constructed in a manner similar to the neon model described in the previous paragraph. For example, for ion stages described by ground configurations with $n=2$ as their valence shell, and with at least two electrons in that $n=2$ shell, i.e., Ar$^{8+}$ -- Ar$^{14+}$, double promotions were considered from the $n=2$ shell up to excited configurations of the form $\ldots 3l\,nl'$, except that the Rydberg electron range is extended to $3 \le n \le 10$ in this case. However, configurations of the form $\ldots 4l\,4l'$ were excluded for these ion stages. For ion stages with ground configurations described by an $n=3$ valence shell, i.e., Ar$^{0+}$ -- Ar$^{7+}$, we consider single-electron permutations out of the $n=2$ or $n=3$ shells. For just the Ar$^{7+}$ stage, which contains a single $n=3$, i.e $3s$, electron in the ground configuration, we consider two-electron permutations in which the $3s$ electron is permuted to a $3p$ or $3d$ subshell, while an $n=2$ electron is simultaneously promoted to a shell with principal quantum number ranging from $3 \le n \le 10$. For the Ar$^{0+}$ -- Ar$^{6+}$ ion stages, which contain at least two electrons in the $n=3$ valence shell, we consider the promotion of two $n=3$ electrons to produce excited configurations containing orbitals up to $\ldots 4l\,nl'$, where $4 \le n \le 10$. Finally, we consider the promotion of $1s$ electrons for all of the argon ion stages. The above electron permutations result in an argon model that contains 11,274 configurations.

\subsection{Radiative power loss}
The radiative cooling rate of a plasma quantifies the rate at which the plasma loses energy by electromagnetic radiation. This process is significant because it impacts the behavior and stability of the plasma by determining its overall temperature and energy balance. The radiation from impurities is usually categorized into three important radiative phenomena: (i) line radiation resulting from bound-bound (BB) transitions, (ii) free-bound (FB) recombination radiation, and (iii) free-free (FF) bremsstrahlung radiation. The total radiative power loss rate is defined as,
\begin{equation}
p_{\text{rad}} = \dfrac{1}{n_e n_i}\sum_Z \left( P^{\text{BB}}_Z + P^{\text{FB}}_Z  + P^{\text{FF}}_Z \right). 
\end{equation}
Here, $P^{\text{BB}}_Z$, $P^{\text{FB}}_Z$, and $P^{\text{FF}}_Z$ are the radiative power loss associated with bound-bound transitions, free-bound transitions, and free-free transitions for a particular charge state ($Z$), respectively. Usually, in the CE approximation, the contribution of the line radiation for a particular charge state can be estimated using the following expression,
\begin{equation} \label{Eq:Cor_BB1}
P^{\text{BB}}_Z = n_e   \sum_{g<j}  n_{g}^Z ~R_{gj}^Z ~\Delta E_{gj}^Z. 
\end{equation}
Here, $g$ and $j$ refer to the ground and excited state indices, respectively. $n_{g}^Z$ and $\Delta E_{gj}^Z$ represent the population density of the ground state and the corresponding transition energy, respectively. $R_{gj}^Z $ is the electron-impact excitation rate from state $g$ to $j$. In the CE model, the population of a particular excited state can be determined by assuming that the electron-impact excitation rate (per unit volume) from the ground state to the excited state equals the spontaneous rate (per unit volume) from a particular excited state to the ground state, i.e., $n_{j}^Z A_{gj}^Z = n_{g}^Z R_{gj}^Z$. Therefore, Eq.~\ref{Eq:Cor_BB1} can be rewritten as,
\begin{equation} \label{Eq:Cor_BB2}
P^{\text{BB}}_Z = n_e   \sum_{g<j}  n_{j}^Z ~A_{gj}^Z ~\Delta E_{gj}^Z. 
\end{equation}

In contrast to the CE approximation, the CR modeling incorporates transitions between different excited states in addition to the ground-to-excited-state transitions. It results in more complex contributions to the line radiation and the resulting bound-bound radiation can be defined as,
\begin{equation}
P^{\text{BB}}_Z = n_e   \sum_{i<j}  n_{j}^Z ~A_{ij}^Z ~\Delta E_{ij}^Z.  
\end{equation}
Here, $n_{j}^Z$ and $\Delta E_{ij}^Z$ represent the population density of the $j^{\text{th}}$ state (upper state) and the corresponding transition energy, respectively. $A_{ij}^Z$ is the spontaneous emission rate from the upper state $j$ to the lower state $i$.
We note that the above sum includes contributions from autoionizing states, which are sometimes included separately in the free-bound contribution in connection with the process of dielectronic recombination.

The free-bound contribution to the radiative power loss is produced during the radiative recombination process.
It is defined as
\begin{equation}
P^{\text{FB}}_Z = n_e    \sum_{i<j}  n_{i}^Z ~\beta_{ij}^Z ~(\chi^{Z-1} + \left\langle E_e \right\rangle ),
\end{equation}
where  $\beta_{ij}^Z$ is the radiative recombination rate.  $\chi^{Z-1}$ and  $\left\langle E_e \right\rangle$ represent the ionization potential and average kinetic energy of recombining electrons, respectively.

Again, we note that the contribution from autoionizing states is not included in the free-bound expression above. Instead, we treat the autoionizing states on par with the bound states, obtaining populations for both types of states by solving the CR equations and then obtaining their line emission via the process of spontaneous decay. Therefore, an explicit contribution from the process of dielectronic recombination does not appear in the above free-bound expression.


Nevertheless, it is sometimes useful to distinguish between the radiative recombination and dielectronic recombination contributions to the radiative power loss. For example, at low temperatures, it is usually the case that direct recombination, i.e., the radiative recombination process, is important. Whereas, for relatively higher temperatures ($T_e \simeq Z^2$ Ry), the resonant recombination, i.e., dielectronic recombination, process dominates as the radiative recombination rates rapidly decrease with increasing temperature \cite{PhysRevE.63.046407}.

For a pure Coulomb field, the resulting radiation from free-free transitions \cite{karzas1961electron}, $P^{\text{FF}}$,  is defined as follows,  

\begin{equation}
P_{\text{FF}} = 9.55 \times 10^{-14} ~ Z_{\text{eff}} ~ n_e ~ g_f,
\end{equation}
where $Z_{\text{eff}}$ is the effective ion charge state and $g_f$ is the free–free Gaunt factor. In this work, $g_f$ is taken from published tabulations of relativistic free–free Gaunt factors. The ATOMIC calculations employ the tables of Nakagawa, Kohyama, and Itoh, which were developed for dense, high-temperature stellar plasmas \cite{Nakagawa1997}, while the FCR calculations use the van Hoof Gaunt factor database, which provides accurate nonrelativistic and relativistic free–free Gaunt factors over an extended parameter range \cite{vanHoof2014,vanHoofDB}.

\begin{figure}[!h]
\centering
\includegraphics[scale=0.26]{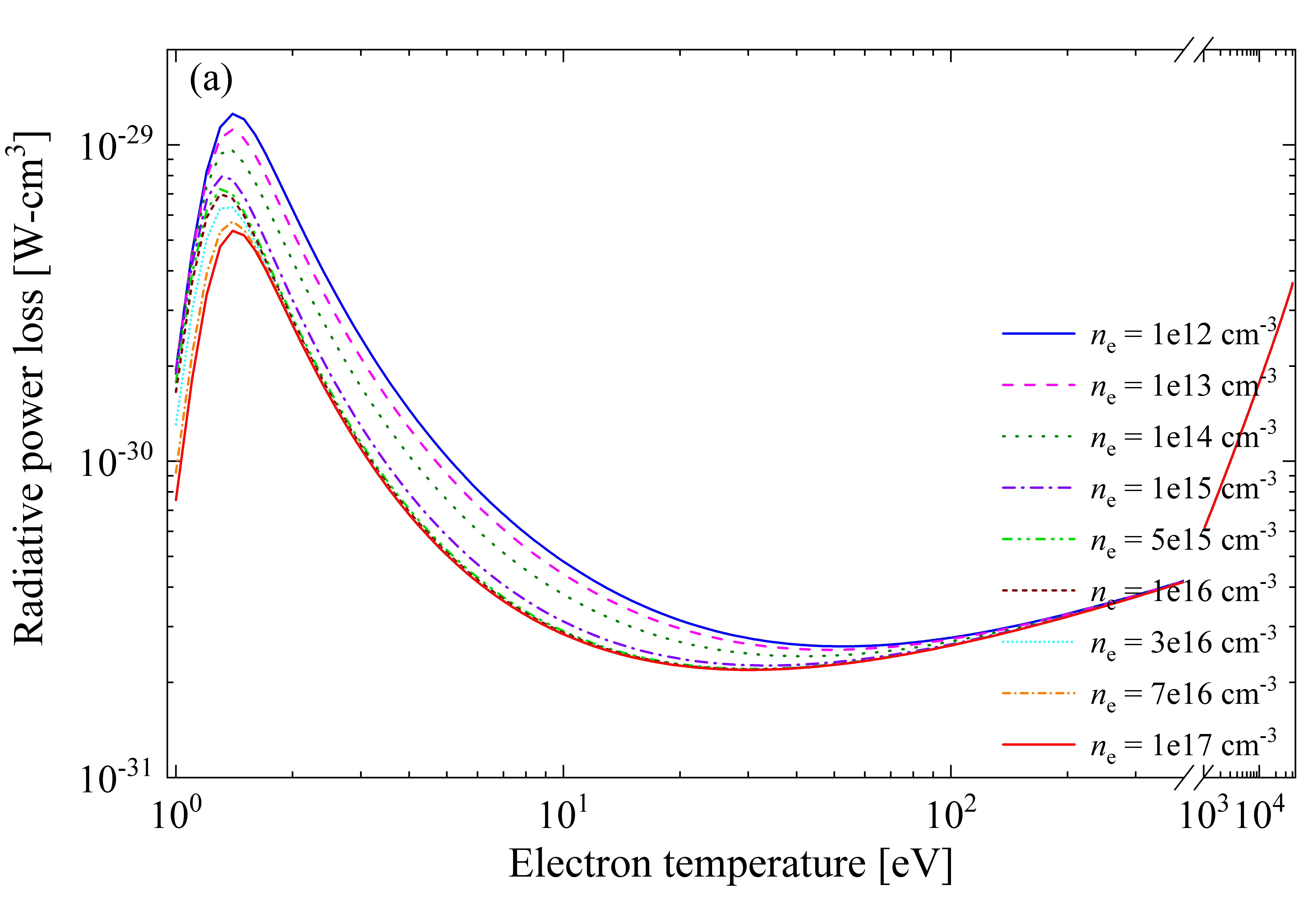}\includegraphics[scale=0.26]{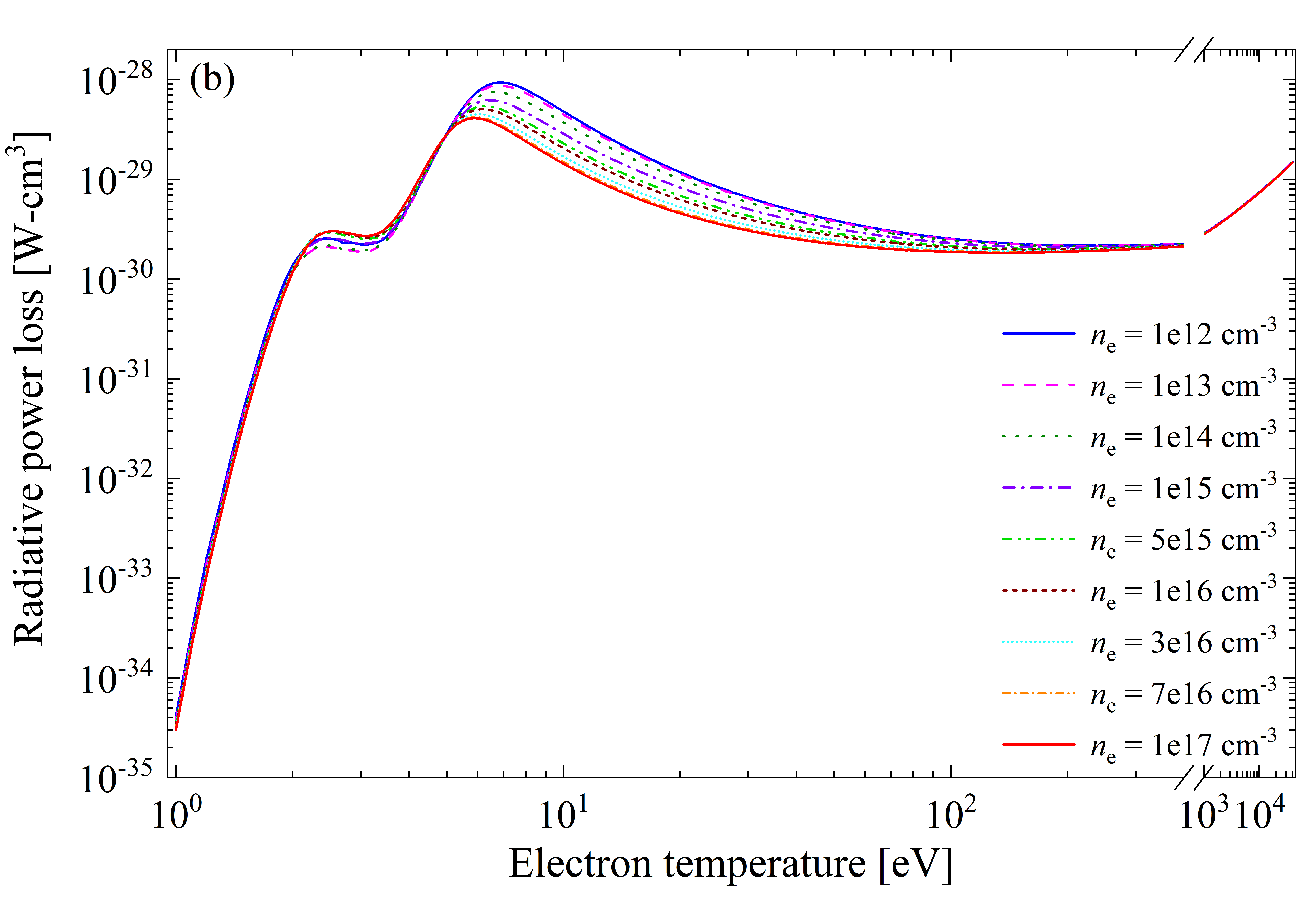}
\includegraphics[scale=0.26]{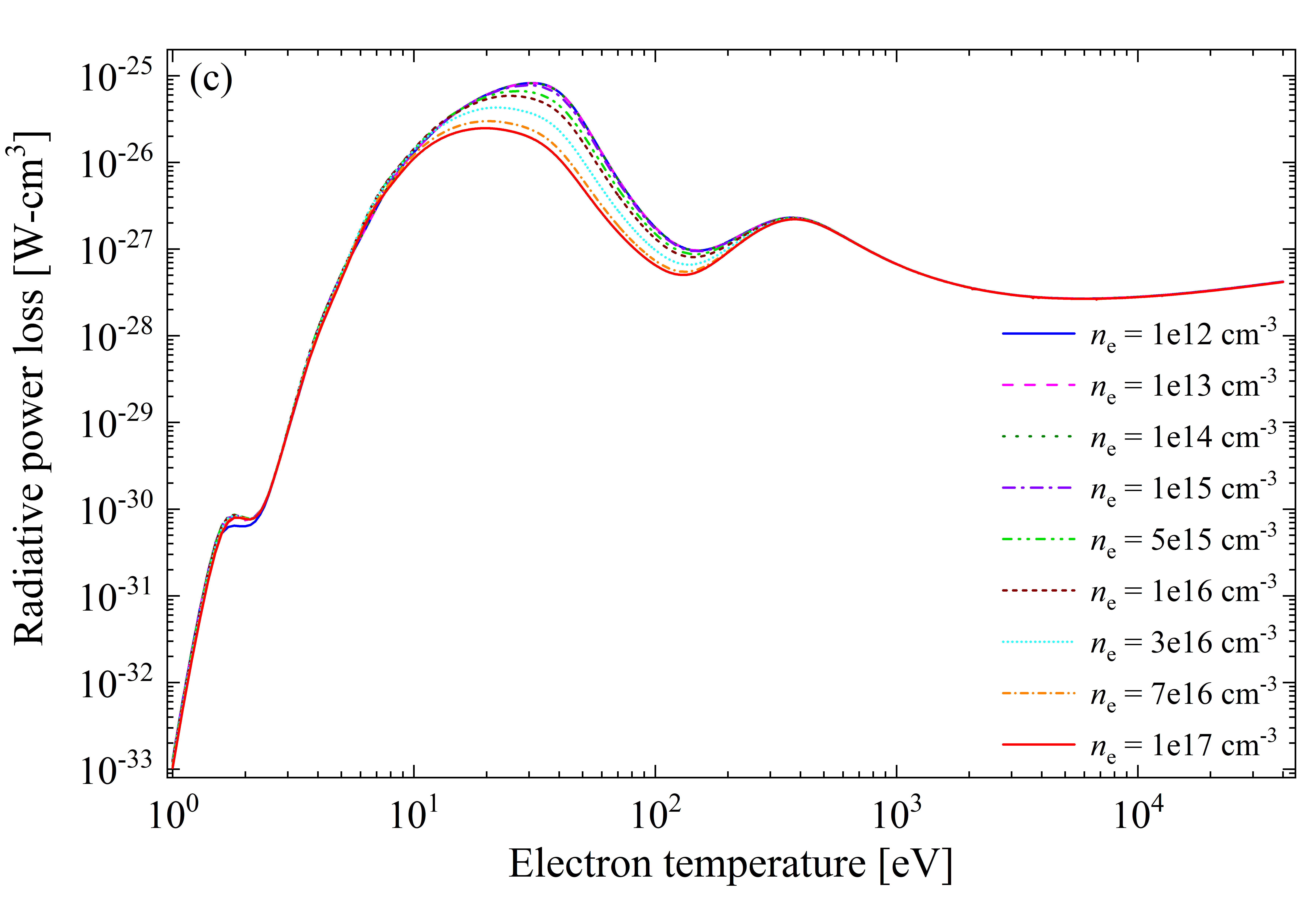}\includegraphics[scale=0.26]{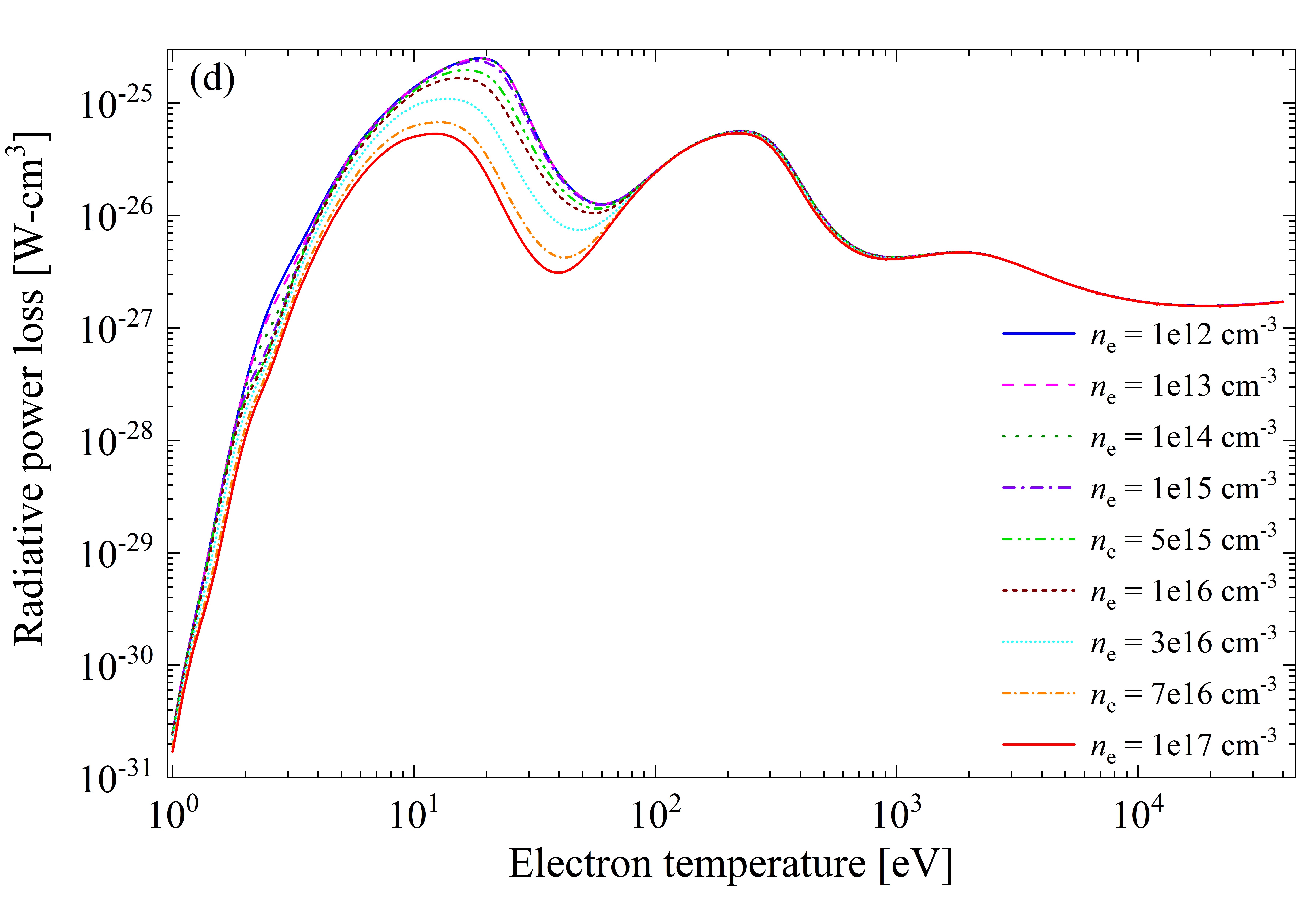}
\caption{\label{fig:prad_atomic} Variation of radiative power loss rate with respect to the electron temperature for (a) hydrogen, (b) helium, (c) neon, and (d) argon. The hydrogen results are calculated using the FCR code \cite{sharma2026hybrid}, while the helium, neon, and argon results are calculated using the ATOMIC code \cite{Fontes2015}. The radiative power loss rates are expressed in W-cm$^3$, where 1 W = 10$^7$ erg-s$^{-1}$.}
\end{figure}

The radiative power loss rate is also affected by the electron and ion densities of the plasma. In general, plasmas with higher densities will have faster cooling rates due to the larger number of particles, which results in more frequent collisional encounters and more possibilities for energy loss by radiation. The evolution of the radiative power loss rate as a function of electron temperature at different electron densities is presented in Fig.~\ref{fig:prad_atomic}(a), \ref{fig:prad_atomic}(b), \ref{fig:prad_atomic}(c), and \ref{fig:prad_atomic}(d) for hydrogen, helium, neon, and argon, respectively.

\begin{figure}[!h]
\centering
\includegraphics[scale=0.18]{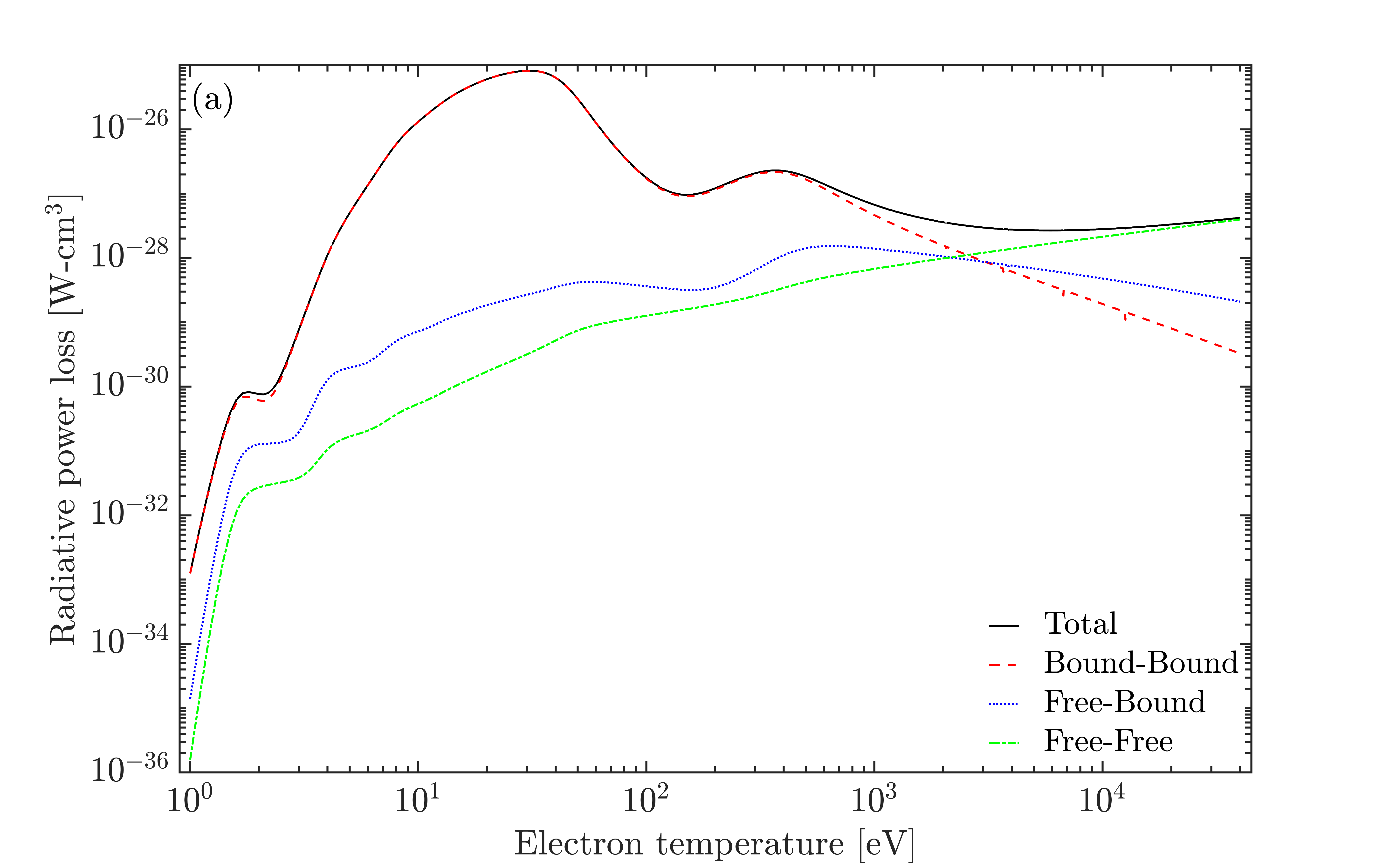}\includegraphics[scale=0.18]{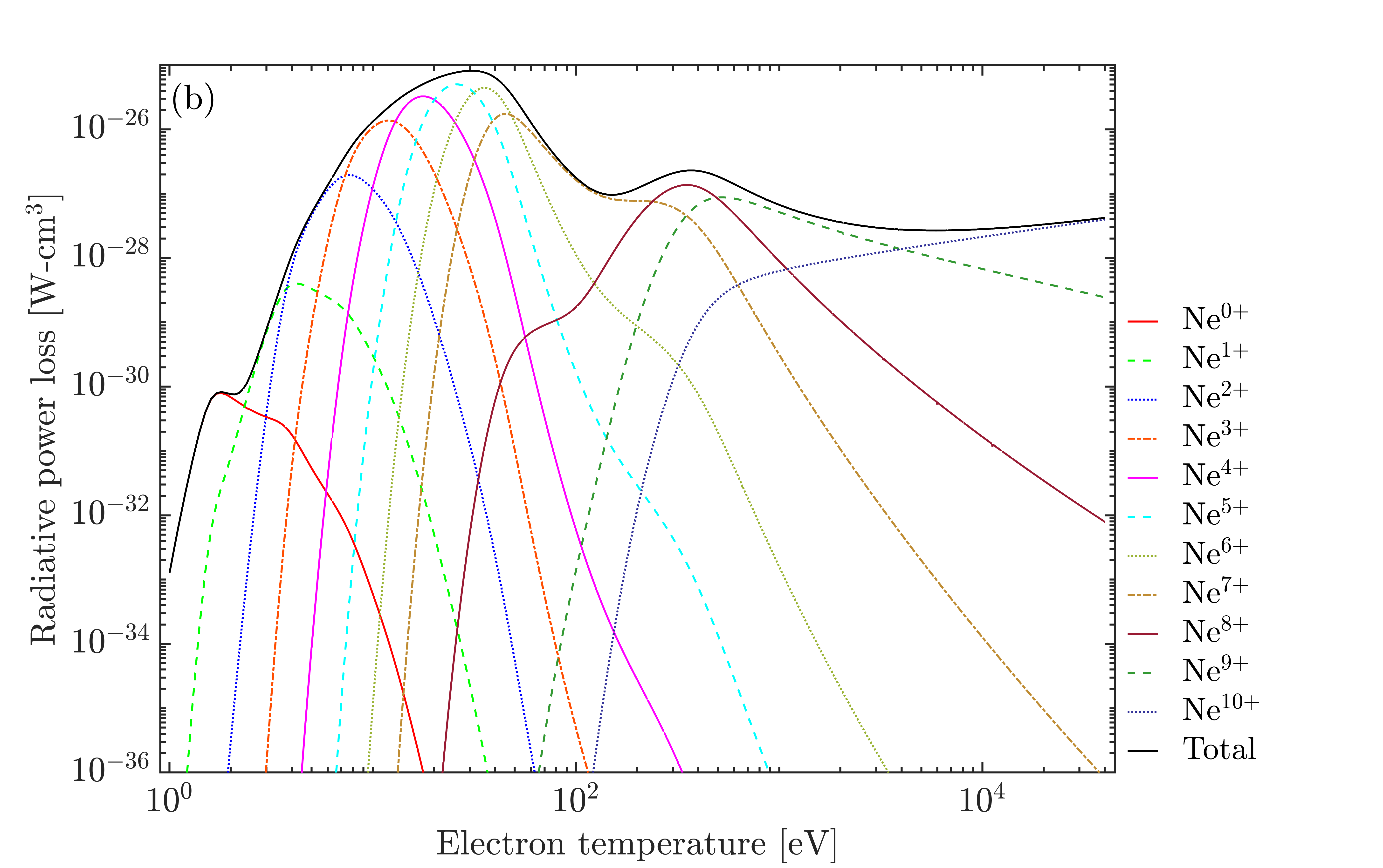}
\caption{\label{Ne_comp_zrad_ind} Radiative power loss calculated for neon as a function of electron temperature for the electron density of $n_e$ = 10$^{14}$ cm$^{-3}$: (a) Contribution of bound-bound, free-bound, and free-free transitions in the total radiative power loss. (b) Contribution of different charge states in the total radiative power loss. All results shown are calculated using the ATOMIC code \cite{Fontes2015}. The radiative power loss rates are expressed in W-cm$^3$, where 1 W = 10$^7$ erg-s$^{-1}$.}
\end{figure}

In Fig.  \ref{Ne_comp_zrad_ind},  we have shown the variation of radiative power loss rate of neon with respect to electron temperature for a particular case of electron density of 10$^{14}$ cm$^{-3}$. The variation represents the characteristic behavior of radiative power loss where the hump-like structure represents the contribution of the different shells (i.e., K and L shells for neon)  of the atom, in Fig.~\ref{Ne_comp_zrad_ind}(a). Further, the contribution of bound-bound, free-bound, and free-free transitions are also depicted in the same figure. It is evident that line radiation resulting from the bound-bound transitions is the most dominant process over the low to intermediate temperature range, whereas bremsstrahlung radiation is crucial only in the higher temperature range where the bound electrons are comparatively fewer. The contribution of free-bound transitions increases with the increase in the atomic number. The contribution of different charge states is presented in Fig.~\ref{Ne_comp_zrad_ind}(b). It is important to note that the peaks corresponding to closed shells are comparatively wider than the other charge state peaks due to the higher binding energies of the electrons in closed shells. 

\subsection{Average charge state}
\begin{figure}[!h]
\centering
\includegraphics[scale=0.26]{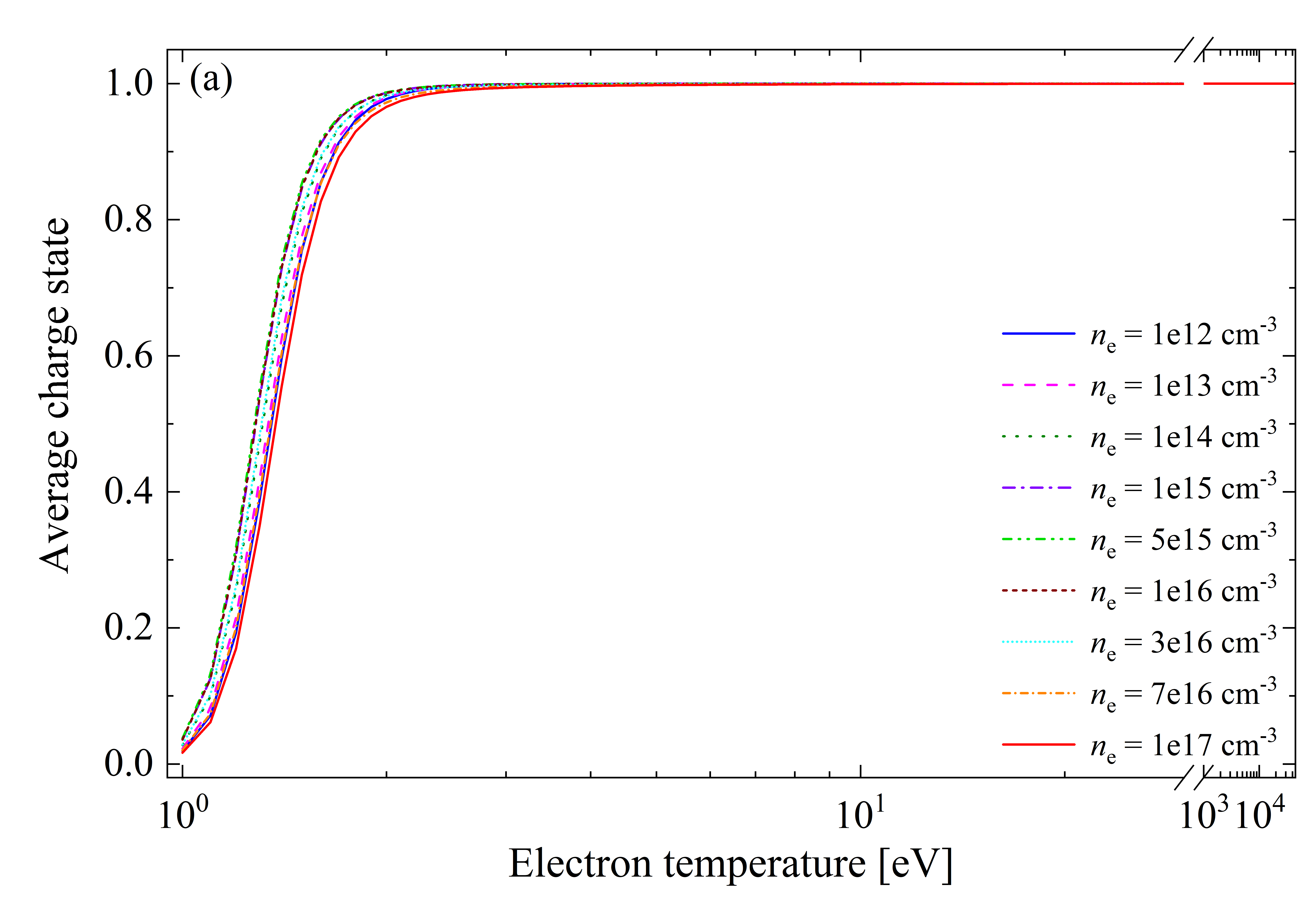}\includegraphics[scale=0.26]{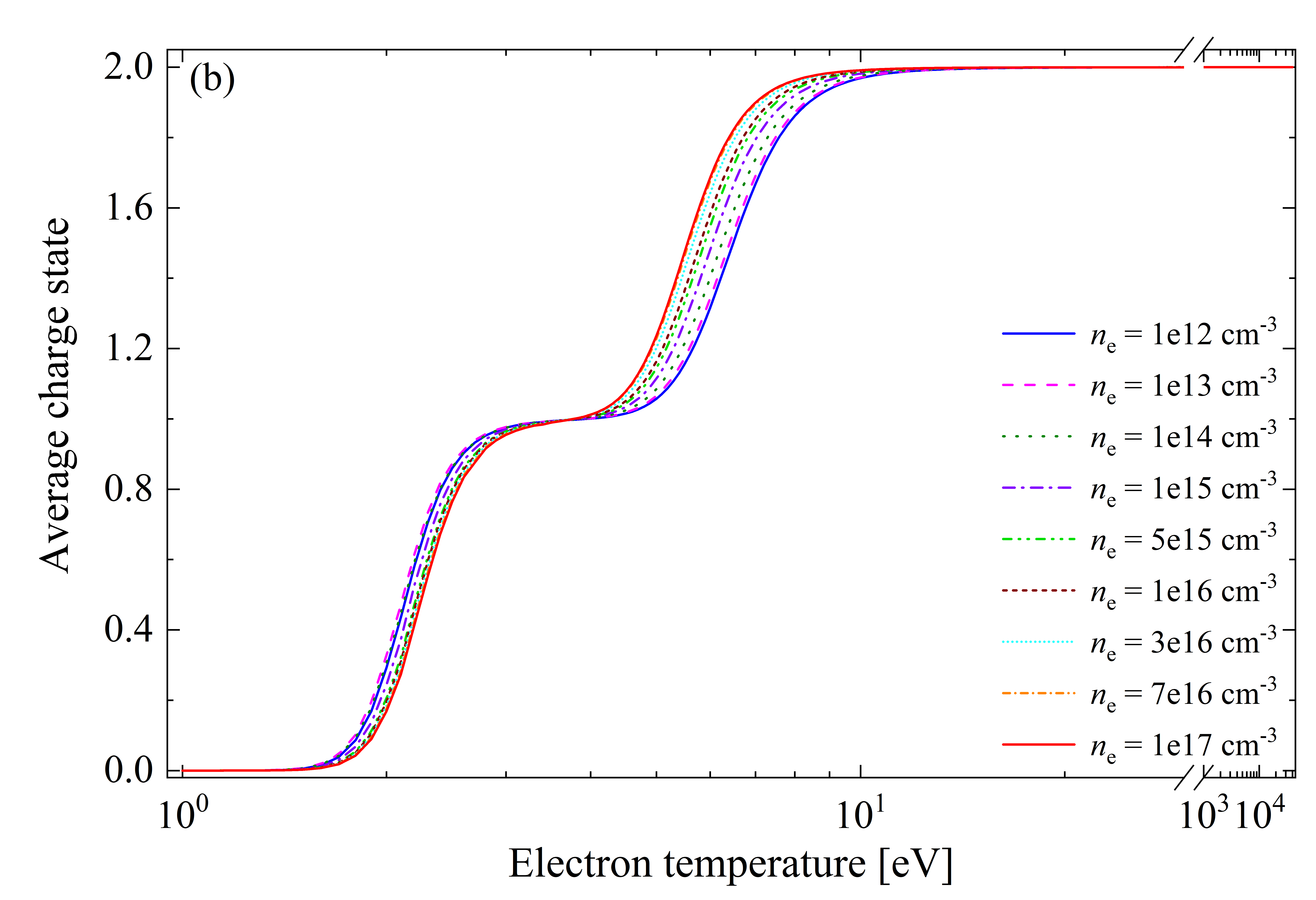}
\includegraphics[scale=0.26]{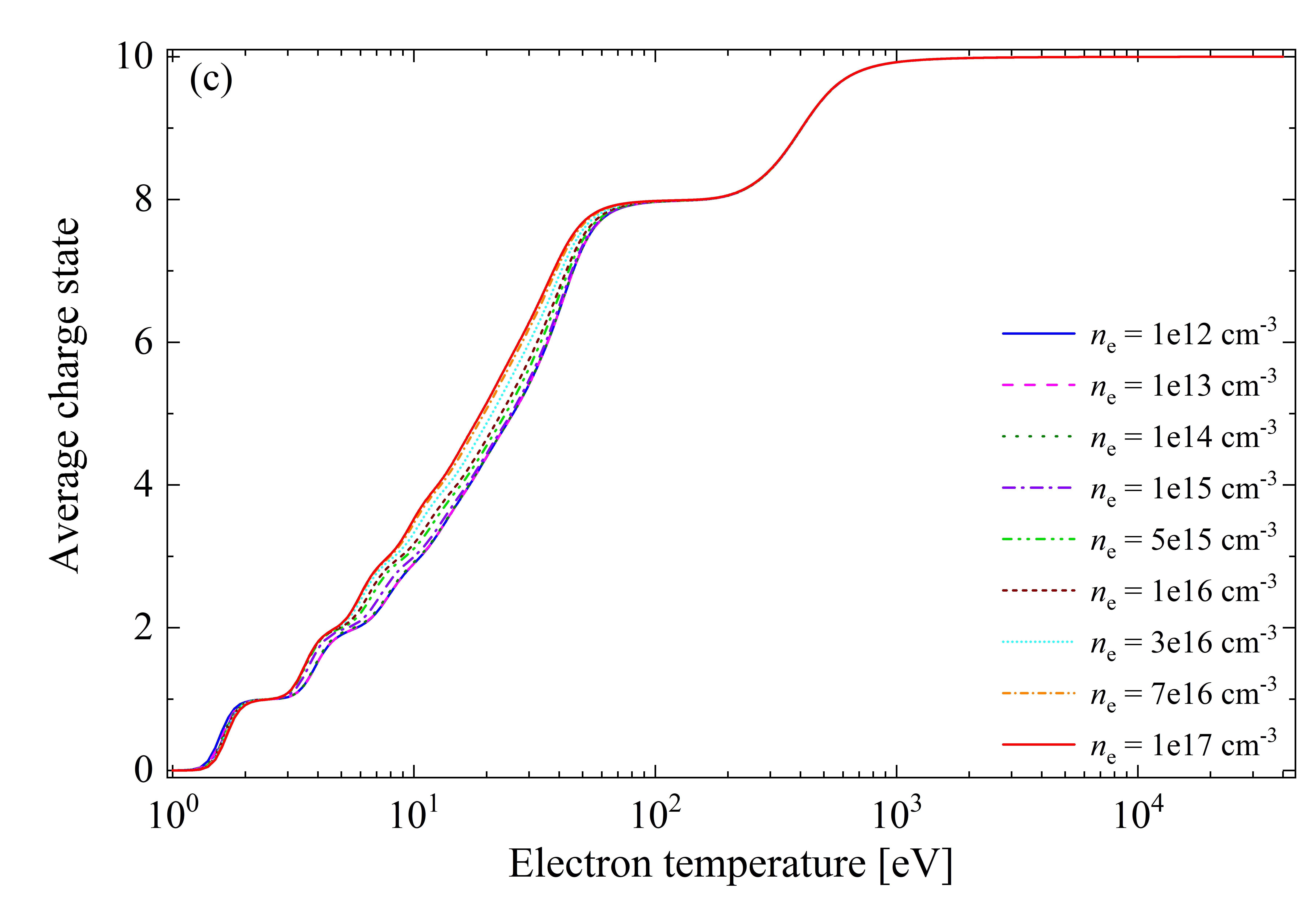}\includegraphics[scale=0.26]{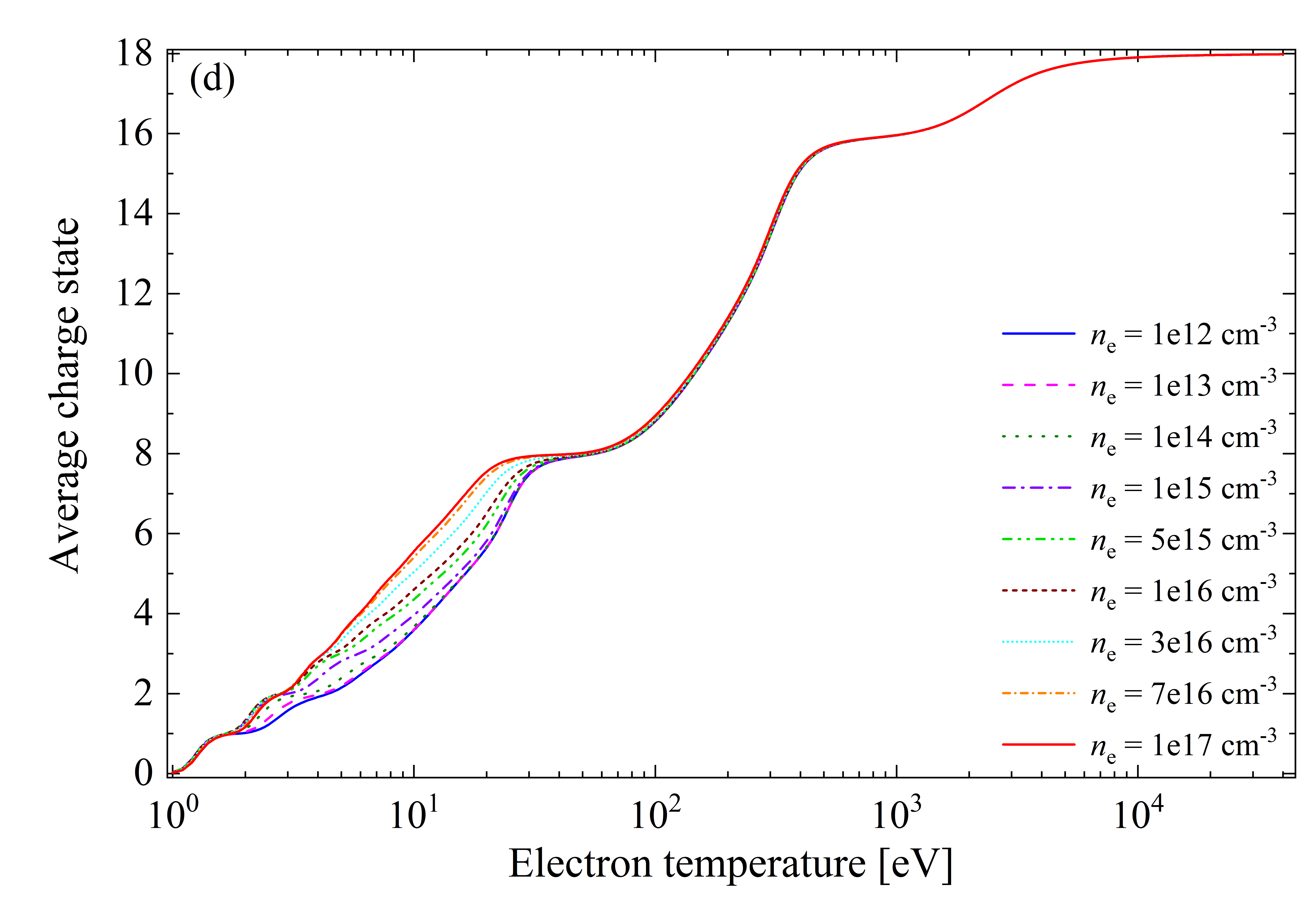}
\caption{\label{fig:zavg_atomic} Variation of average charge state with respect to the electron temperature for (a) hydrogen, (b) helium, (c) neon, and (d) argon. The hydrogen results are calculated using the FCR code \cite{sharma2026hybrid}, while the helium, neon, and argon results are calculated using the ATOMIC code \cite{Fontes2015}.}

\end{figure}

The average charge state of the plasma species can also be expressed in terms of the charge state fraction ($f_Z$) of the species, such that, 
\begin{equation}
Z_{\text{avg}}  = \sum_Z  Z ~ f_Z,
\end{equation}
where $f_Z = n^Z/n_i$ is the fractional population of charge state $Z$. The variation of average charge state with respect to electron temperature for hydrogen, helium,  neon,  and argon is shown in Fig.  \ref{fig:zavg_atomic}(a),  \ref{fig:zavg_atomic}(b),  \ref{fig:zavg_atomic}(c), and \ref{fig:zavg_atomic}(d), respectively. It is observed that the average charge state of the plasma increases with the increase in electron temperature, which can be attributed to the greater kinetic energy that electrons acquire at higher temperatures, allowing them to overcome the ionization potentials of the atoms/ions in the plasma. The intermediate plateaus in the curves correspond to closed-shell electronic configurations, where the significantly higher ionization potential requires substantially elevated electron temperatures to achieve further ionization. For higher electron densities ($n_e \geq 10^{15}$ cm$^{-3}$), a significant deviation has been observed in the average charge state values. With the increase in electron density,  the probability for electron impact excitation increases, and consequently, the population in excited states and metastable states also increases. As a result,  the various electron transitions among excited and metastable states become important with the change in electron temperature and electron density. It is noteworthy that being closer to the continuum, the metastable states often have higher ionization rates than the ground state. It clearly suggests that in typical plasma conditions,  as the electron density increases,  the effective ionization of the atomic species also increases. Further, at quite high electron density, the three-body recombination process, which is proportional to the square of electron density,  also plays an important role in the ionization balance. 

\subsection{Effective charge state}
\begin{figure}[!h]
\centering
\includegraphics[scale=0.26]{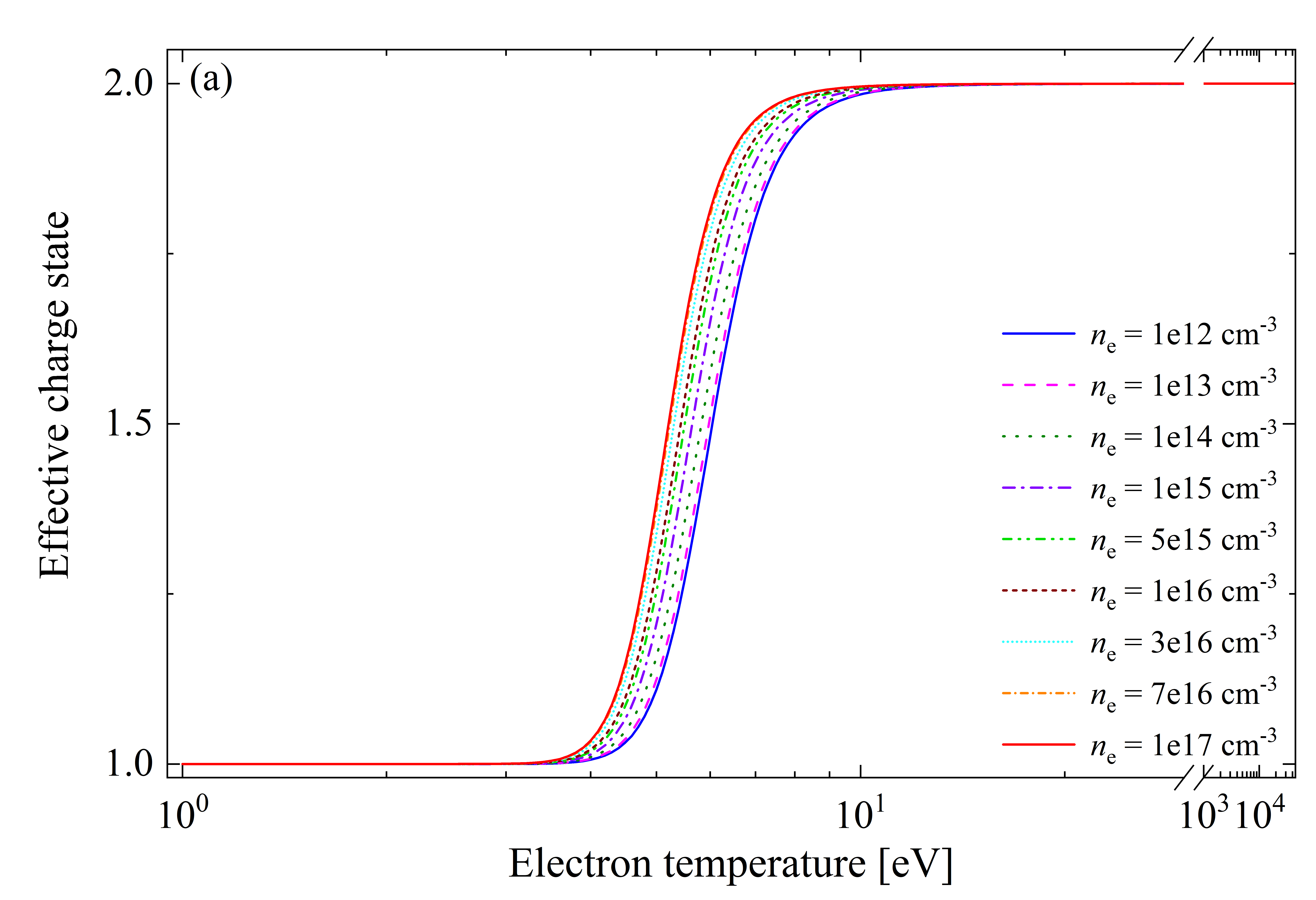}\includegraphics[scale=0.26]{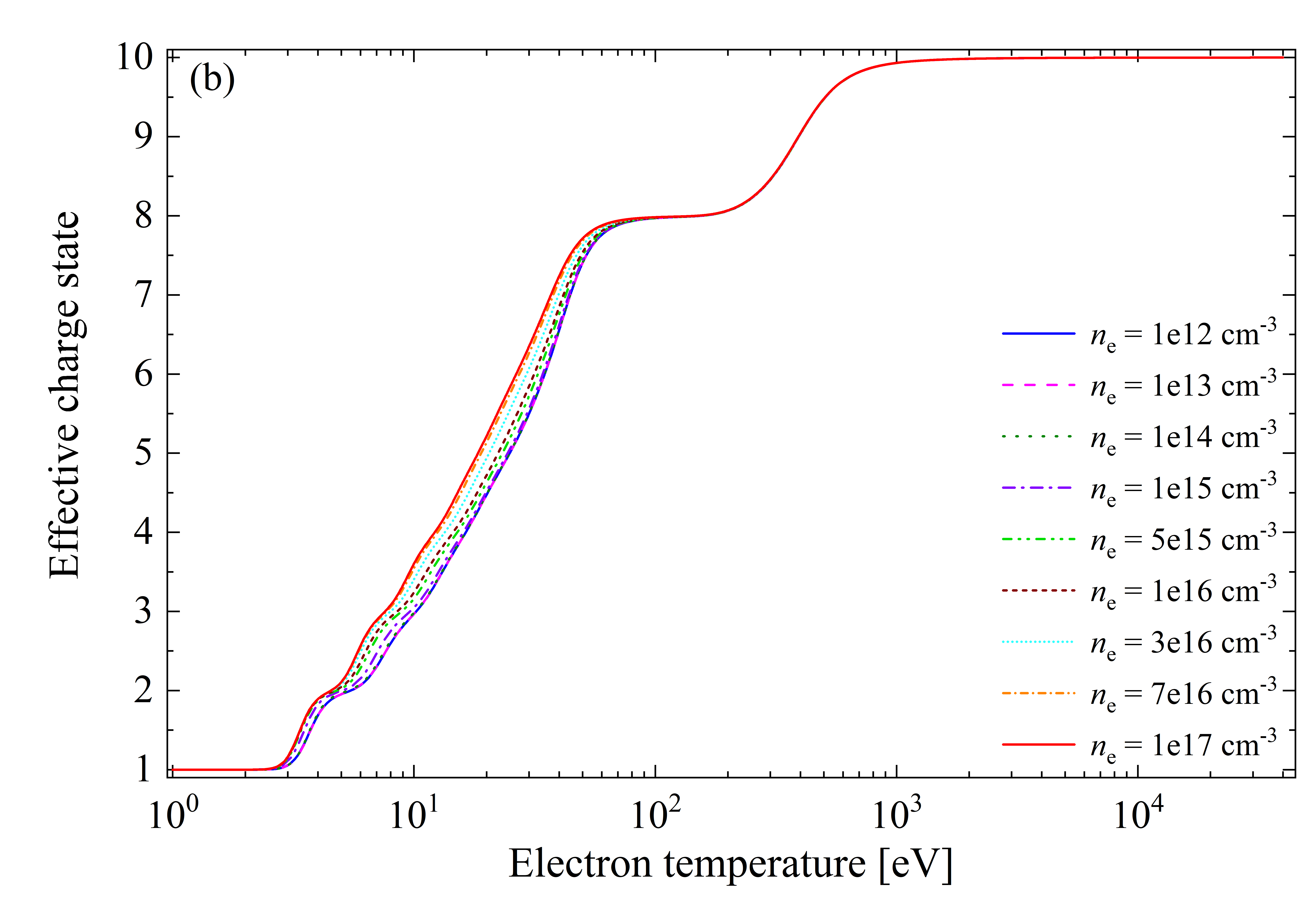}
\includegraphics[scale=0.26]{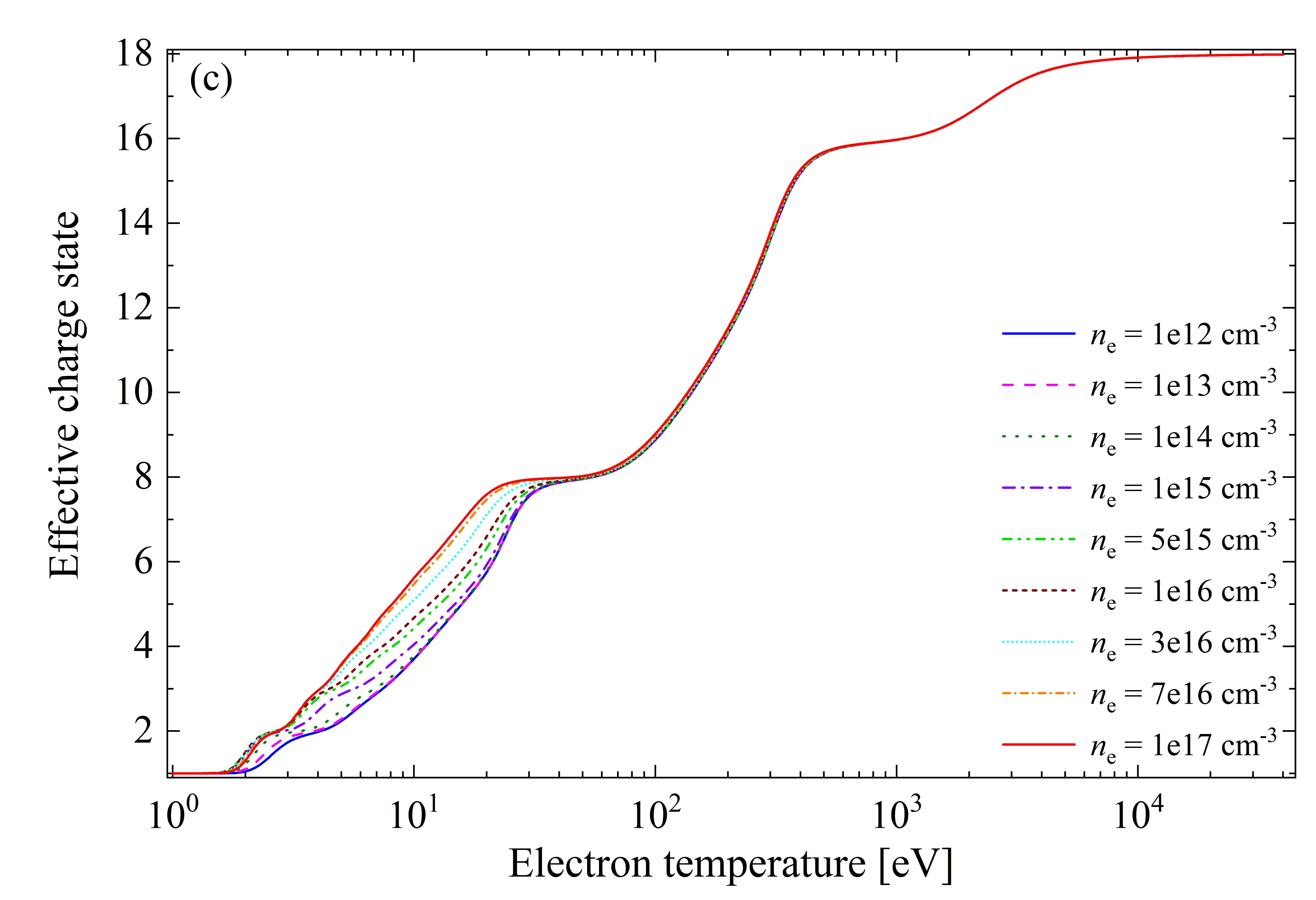}
\caption{\label{fig:zeff_atomic} Variation of effective charge state with respect to the electron temperature for (a) helium, (b) neon, and (c) argon. All results shown are calculated using the ATOMIC code \cite{Fontes2015}.}
\end{figure}

The effective charge state is considered one of the key plasma parameters to characterize the impurity concentration in the plasmas\cite{kadota1980space,rathgeber2010estimation}, electron-ion/atom conductivity\cite{hirshman1978neoclassical}, and energy loss of heavy ions in plasmas\cite{deutsch2016ion,ren2023target}. In terms of charge state fraction, it is defined as,
\begin{align}
Z_{\text{eff}} = \dfrac{\sum_{Z} Z^2 f_Z}{Z_{\text{avg}} }.
\end{align}
In Fig.~\ref{fig:zeff_atomic}(a),  \ref{fig:zeff_atomic}(b), and \ref{fig:zeff_atomic}(c),  we have presented the evolution of the effective charge state with respect to the electron temperature for helium, neon, and argon,  respectively.

\section{Comparisons with superconfiguration CE model and  CR model\label{sec:model_compare}}

\subsection{Model reduction with coronal equilibrium approximations}
The CE approximation, also known as collisional-ionization equilibrium approximation, is a limiting case, where the radiative processes are prominent compared to the collision-based ones. This approximation is relevant in situations with low-density (optically thin) and high-temperature plasmas, such as the solar corona, other stellar coronae, and interstellar space. The model was initially developed for understanding the solar corona but has since been adapted for other astrophysical plasmas and more recently became quite popular in tokamak disruption modeling \cite{Whyte:2002}. A significant amount of effort is dedicated to ongoing research and development of codes or atomic databases for conducting spectral analysis and calculating radiative cooling rates of astrophysical plasmas. There are three commonly used plasma codes, i.e., SPEX \cite{kaastra2018spexx}, AtomDB/APEC \cite{smith2001collisional,foster2012updated,foster2020pyatomdb}, and CHIANTI \cite{dere1997chianti,del2021chianti}, utilized for analyzing astrophysical spectra and calculating radiative cooling rates using the CE model. In the CE approximation, the electron impact excitation rate is relatively low compared to the spontaneous emission rate due to the condition of low electron density, and therefore, the population densities of all the excited states are comparatively low. Consequently, the ions can be assumed to be in their ground state during collisions (ground state approximation), and secondary collisions with excited states can be disregarded. In this scenario, an ionization balance is achieved when collisional ionization is counteracted by radiative recombination at a given electron temperature. Consequently, density effects are strongly suppressed in the coronal approximation, and key plasma parameters become effectively independent of electron density. As the electron temperature increases, additional atomic processes, such as dielectronic recombination and its inverse process, i.e., excitation-autoionization, begin to influence the ionization balance \cite{burgess1964delectronic,jacobs1977influence,hahn1985theory}. 

\begin{figure}[!h]
\centering
\includegraphics[scale=0.26]{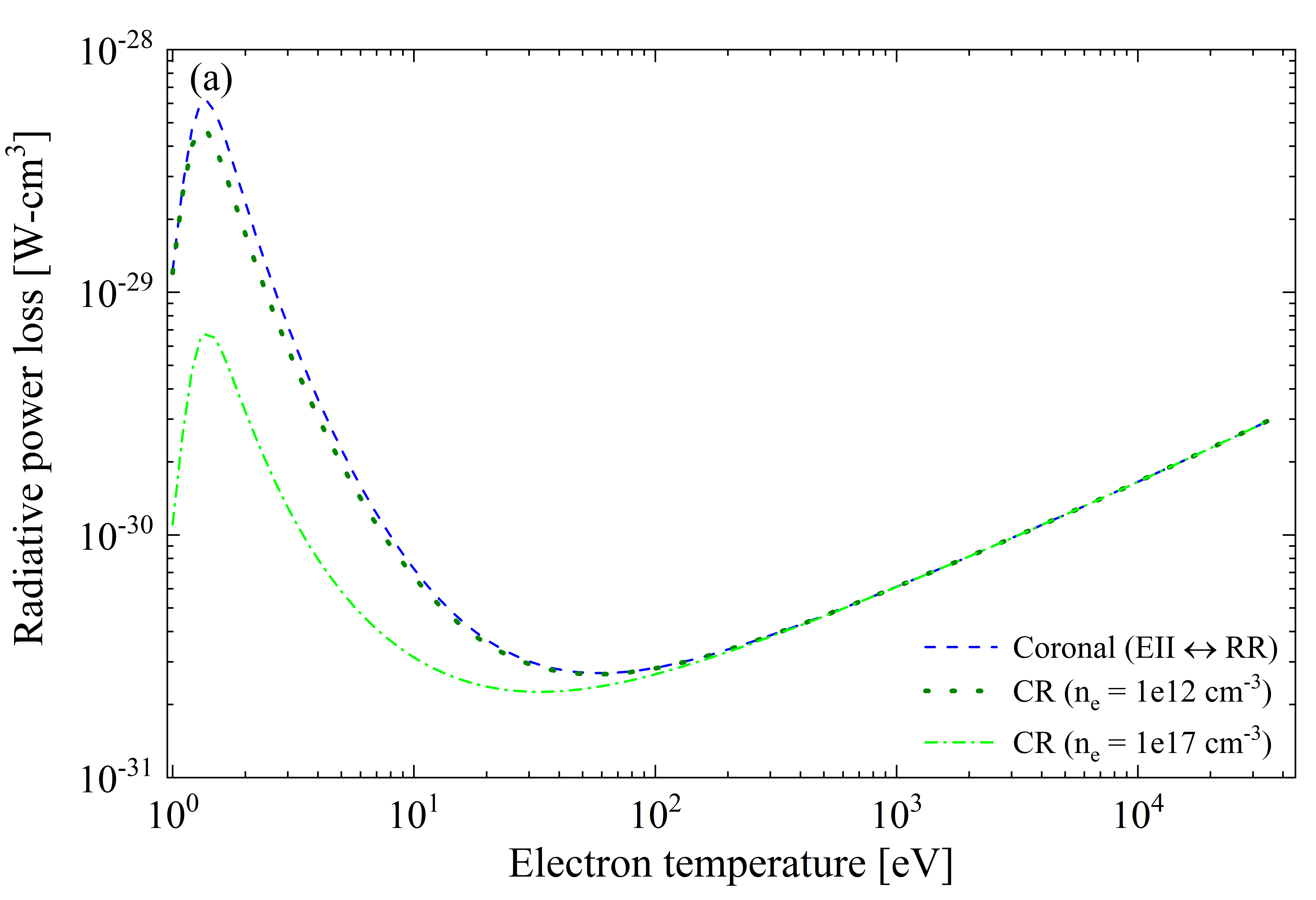}\includegraphics[scale=0.26]{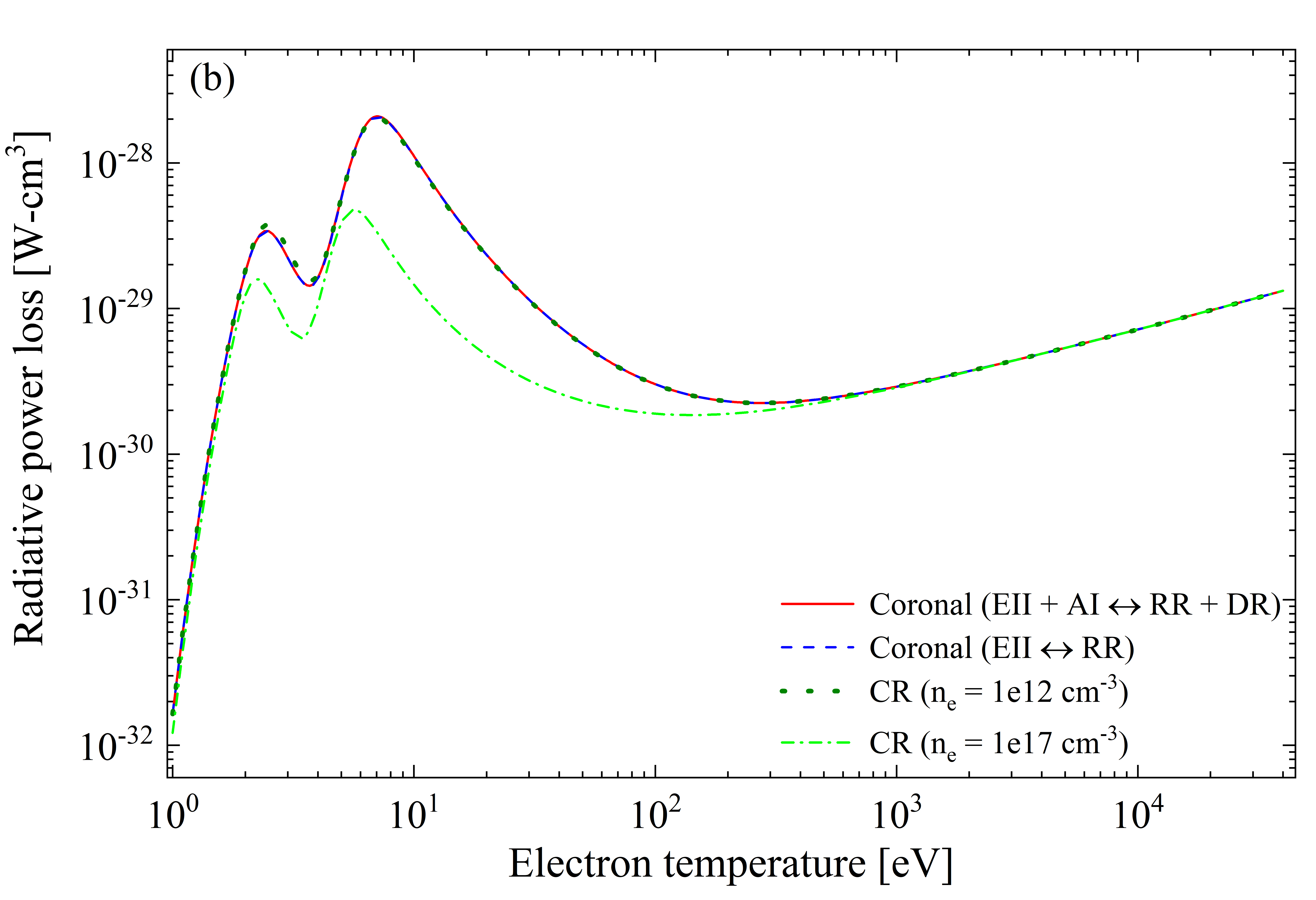}
\includegraphics[scale=0.26]{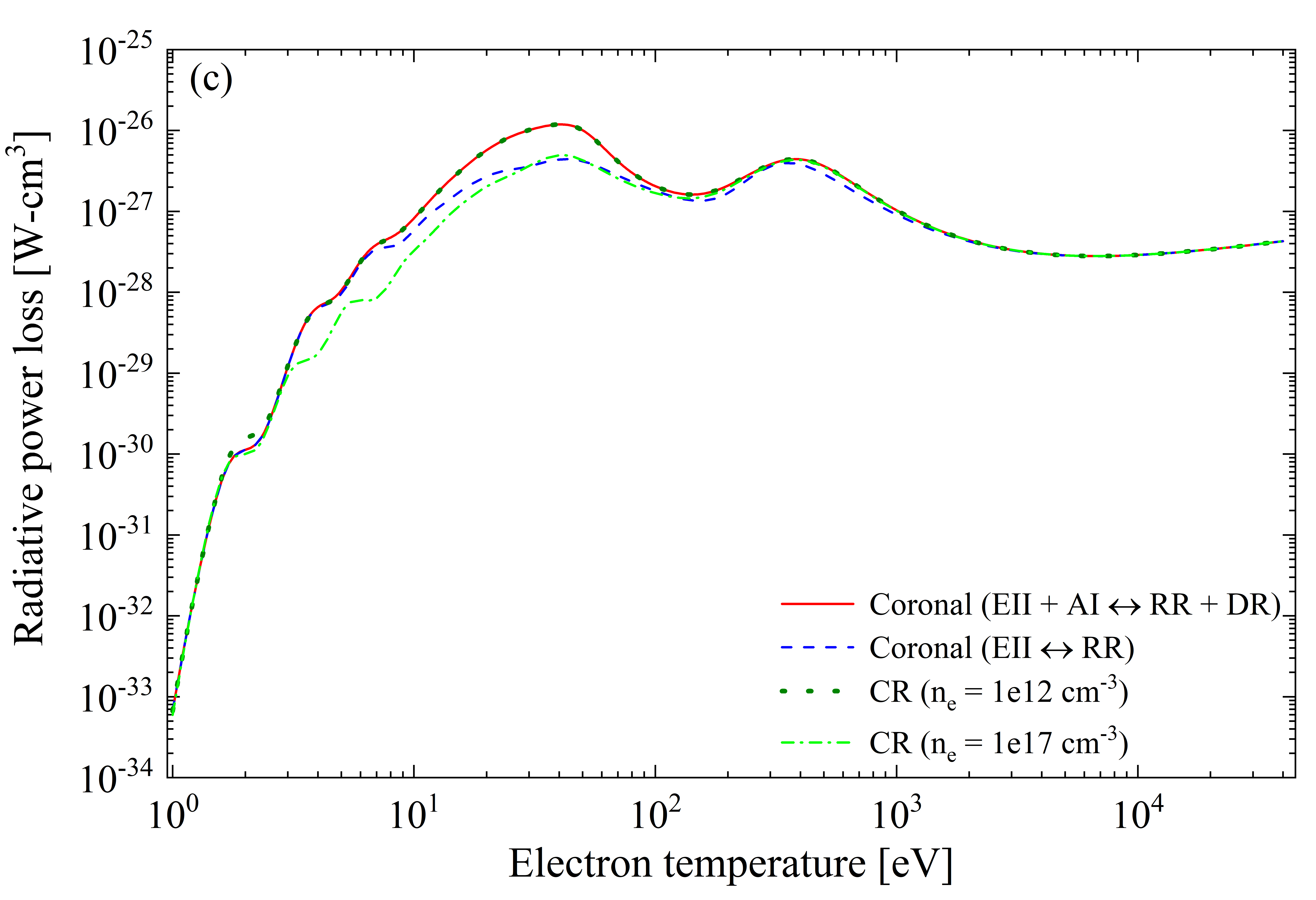}\includegraphics[scale=0.26]{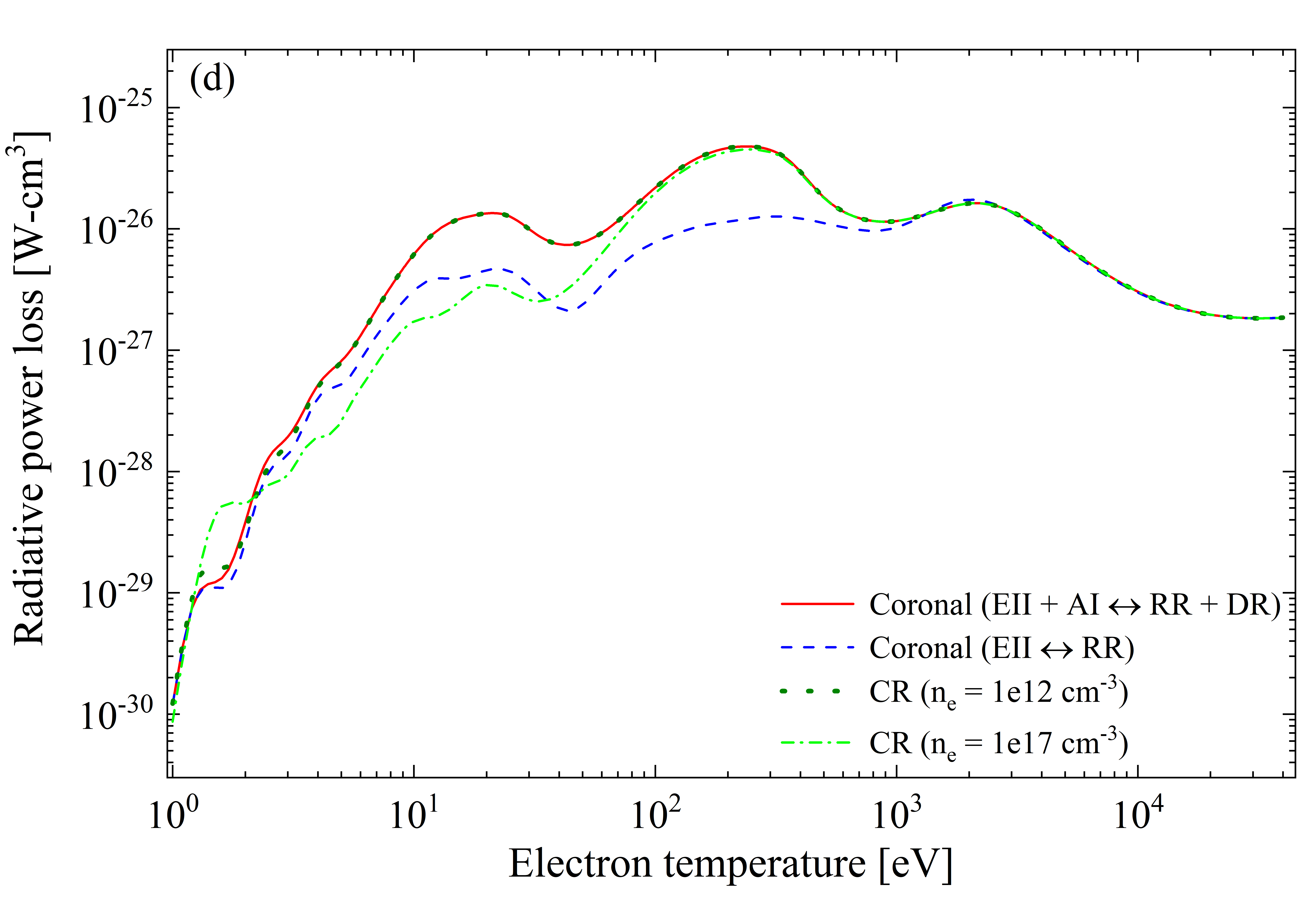}
\caption{\label{Ne_coronal} Variation of the radiative power loss rate as a function of electron temperature for (a) hydrogen, (b) helium, (c) neon, and (d) argon. All results are calculated within the FCR framework using a superconfiguration-based atomic-state representation. The figure compares three models: a full coronal model, a basic coronal model, and a full collisional-radiative model incorporating all relevant atomic processes and their inverse reactions. The definitions of the coronal limits are described in the text. The radiative power loss rates are expressed in W-cm$^3$, where 1 W = $10^{7}$ erg s$^{-1}$.}
\end{figure}

In Fig.~\ref{Ne_coronal}, we show the variation of radiative power loss as a function of electron temperature at different electron densities, with all results calculated within the FCR framework using a superconfiguration-based atomic-state representation. Three distinct models are considered: the first is the full coronal model that achieves equilibrium between different states through the balance among the electron-impact ionization, excitation-autoionization, radiative recombination, and dielectronic capture processes. The second model, i.e., the basic coronal model, maintains the balance between electron-impact ionization and radiative recombination processes only. In both these models, only the electron-impact excitations from the ground to other excited states are considered, and all other excitations are neglected. All the electron-impact de-excitation transitions are neglected. Lastly, the full CR model incorporating all the atomic processes along with their inverse reactions is used to perform the calculations. As discussed earlier, one can observe that the radiative power loss curves are density-independent for both the coronal models. The disparity in radiative power loss derived from these models grows as the number of possible transitions of autoionization and electron capture increases in atomic systems with higher atomic numbers. It is noteworthy that, for lower electron densities, the full coronal model closely approximates the full CR  model, as discussed earlier. Conversely, the comparison contrasts markedly at higher electron densities, where there is a significant divergence between the models. This discrepancy is primarily attributed to the presence of non-radiative de-excitation channels,  such as three-body recombination, electron impact de-excitation, etc.,  at these higher electron densities.

It is also noteworthy that in this simplified approach, i.e., coronal approximation, the contributions from metastable states and transitions between different excited states are generally neglected, and the transitions between ground and excited states are only considered. Interestingly, in various studies \cite{summers2006ionization,ralchenko2016modern,johnson2020effect}, it is noted that metastable states can significantly influence the population dynamics of plasma species. As CR models are developed to encompass the entire spectrum of plasma dynamics, they integrate the metastable states alongside the ground and excited states in the calculations. This ensures accurate predictions of atomic species population distributions compared to the coronal models in intermediate and higher electron density regimes. The inclusion of metastable states also necessitates the inclusion of complex atomic processes, including resonant excitation \cite{PhysRevA.53.3110}, resonant ionization \cite{PhysRevA.92.063430}, and ladder ionization \cite{hansen2007hybrid}. Further, three-body recombination is also expected to play an important role at higher electron densities. Integrating these atomic processes into modeling results in strong density dependence on the plasma parameters as depicted in Figs.~\ref{fig:prad_atomic}, \ref{fig:zavg_atomic}, and \ref{fig:zeff_atomic}, where we have shown the results obtained from CR modeling.

\subsection{CR model reduction with superconfiguration averaging}
Another set of CR model simplifications arises from approximating averaged atomic states, of which the superconfiguration CR model is the coarsest. In this study, we employed the FLYCHK code \cite{CHUNG20053}, which utilizes the superconfiguration method, for comparison with results obtained from the ATOMIC code. The primary objective of a superconfiguration collisional-radiative model is to provide rapid and accurate estimates of the plasma parameters. Interestingly, in certain situations, such as high-temperature plasmas where the contribution of $\Delta n = 0$ transitions to the radiation is negligible, superconfiguration collisional-radiative models can yield satisfactory results.

\begin{figure}[!h]
\centering
\includegraphics[scale=0.18]{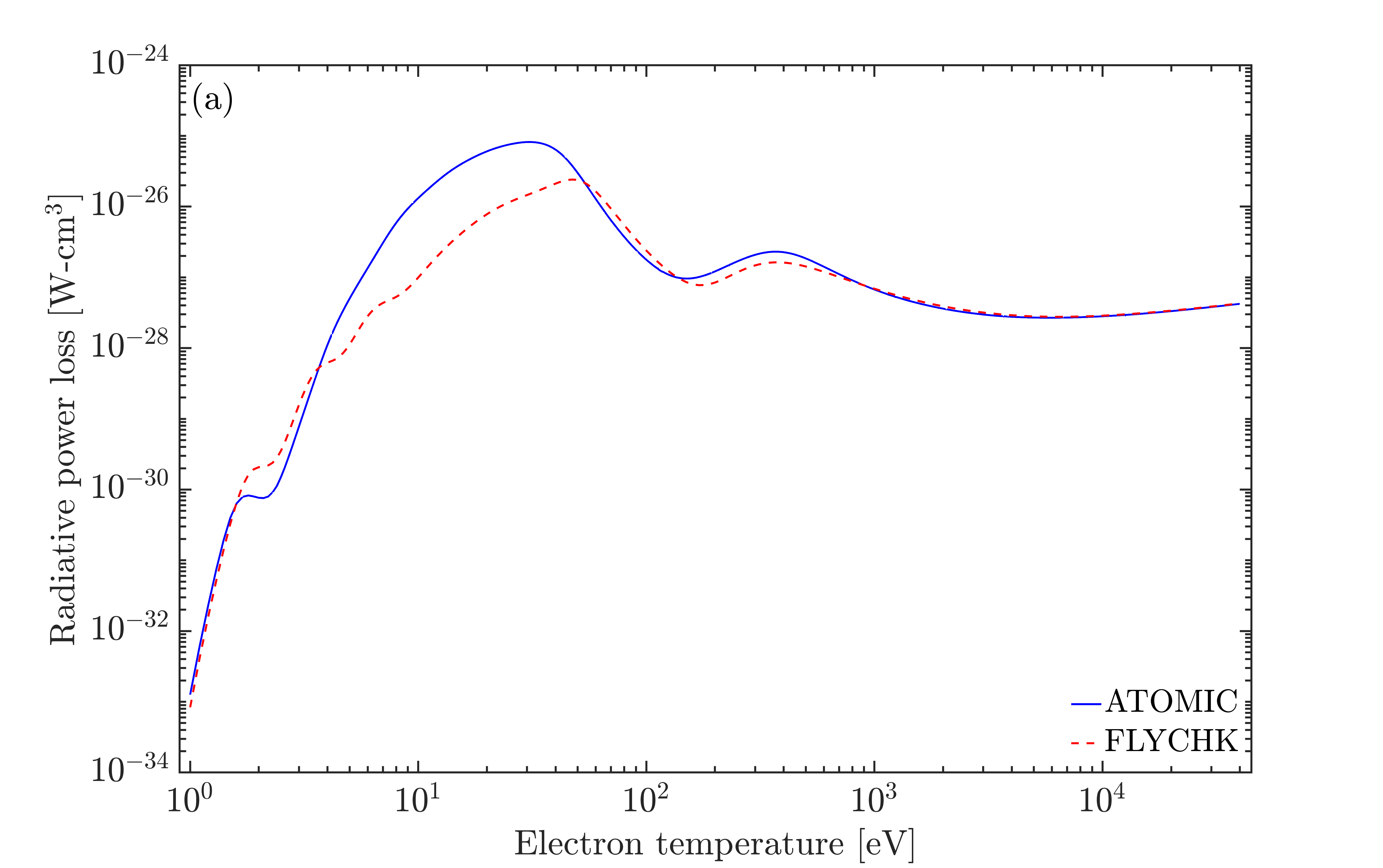}\includegraphics[scale=0.18]{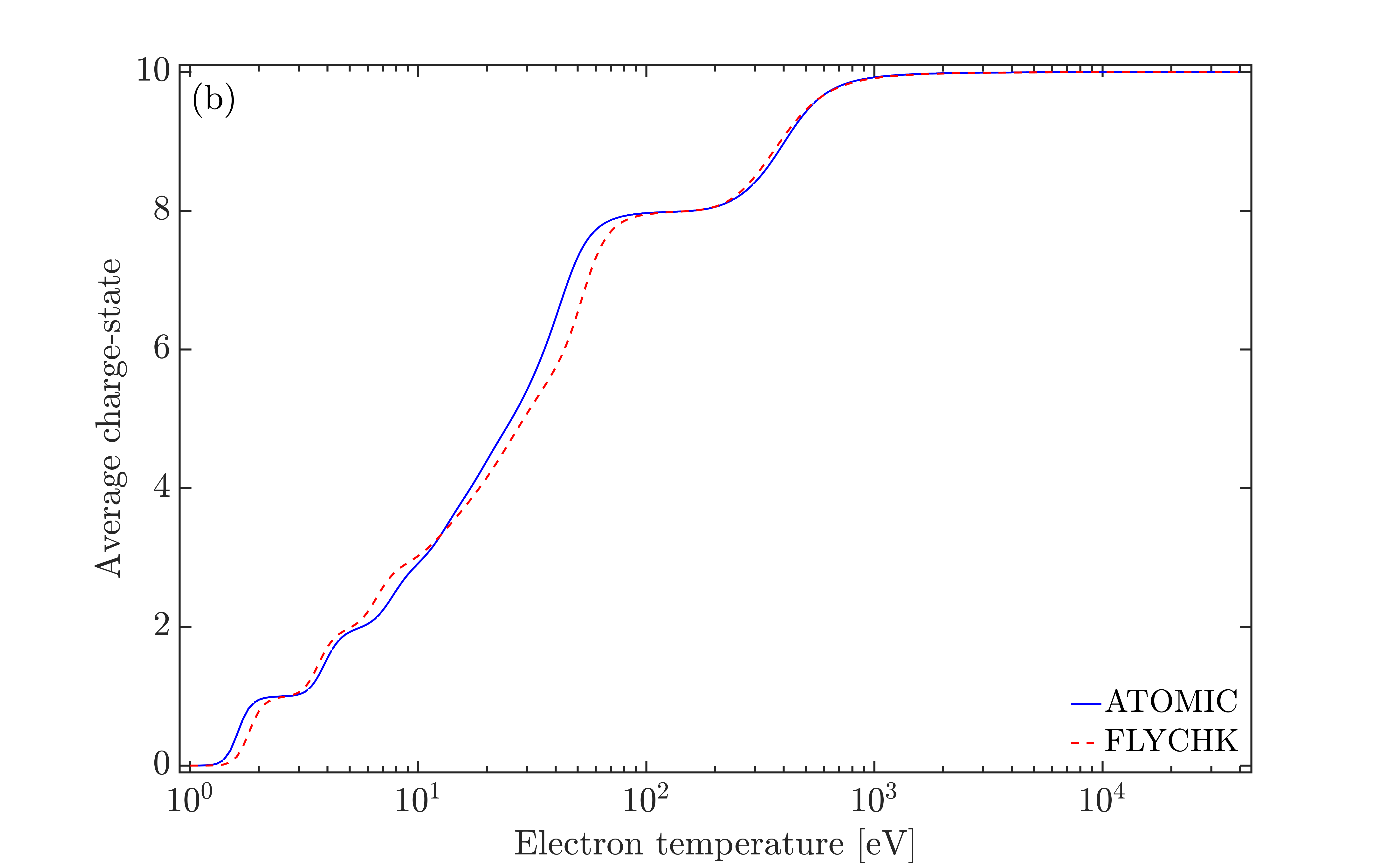}
\caption{\label{fig:Ne_Flychk_Atomic} Comparison of (a) radiative power loss and (b) average charge state calculated using the ATOMIC and FLYCHK codes for a particular case of neon at $n_e = 10^{14}$ cm$^{-3}$. The ATOMIC model used here is configuration-average, and the FLYCHK model \cite{CHUNG20053} uses superconfigurations.}
\end{figure}

Numerous studies \cite{summers1979radiative, badnell2001dielectronic,  vstofanova2021new, jacobs1979dielectronic, breton1978ionization,  jacobs1978ionization, cox1969ionization} have shown that dielectronic recombination and autoionization processes significantly influence plasma parameters. However, a major limitation of the superconfiguration model is that it does not incorporate a comprehensive and essential formulation of these complex atomic processes, relying instead on approximations. Additionally, the superconfiguration model does not account for $\Delta n = 0$ transitions, which can result in considerable errors in radiative power loss calculations when contributions from these transitions are dominant. Both of these shortcomings lead to significant deviations in the values of radiative power loss, as demonstrated in Fig.~\ref{fig:Ne_Flychk_Atomic}, where we have compared the ATOMIC (in its configuration-average mode) results with the FLYCHK (superconfiguration) results for neon for a particular case of $n_e = 10^{14}$~cm$^{-3}$.

\section{Tensor-Product B-spline Surface Fitting}\label{sec:cubic}

As discussed earlier, the accuracy of computed plasma parameters depends on the underlying model and its level of refinement. In practical applications, such as inline use within radiation-hydrodynamics and plasma transport codes, it is often not feasible to repeatedly evaluate computationally expensive CR models. A common alternative is to precompute the CR quantities on a prescribed grid in $(T_e,n_e)$ and then use a surrogate representation that can be evaluated rapidly while retaining the fidelity required for coupled plasma simulations. Previous parametrization studies have typically focused on coronal-limit regimes \cite{post1977steady,summers1979radiative,morozov2007impurity,breton1978ionization,hulse1983numerical}, where plasma parameters are independent of electron density, or on limited sets of plasma species that do not cover the full range required for tokamak disruption mitigation \cite{gil2013parametrization,fournier1998ionization,espinosa2018parametrization,rodriguez2014parametrization}.

In the present work, the CR data are represented using a smooth tensor-product B-spline surface defined over the two-dimensional parameter space of electron temperature and electron density. Instead of performing pointwise bicubic interpolation on a fixed grid, we determine a compact global set of B-spline coefficients that reproduces the tabulated data with high accuracy while preserving spline smoothness. Compared with simpler interpolation schemes, such as nearest-neighbor or piecewise linear methods, the B-spline representation is better suited for integration into coupled plasma models and optimization workflows. It provides controlled smoothness and continuous derivatives determined by the spline degree and knot multiplicities, which is advantageous in multi-dimensional applications.

For each plasma quantity of interest (namely $Z_{\text{avg}}$, $Z_{\text{eff}}$, and $p_{\text{rad}}$), we construct a bivariate spline representation in the transformed variables $x=\log_{10}(n_e)$ and $y=\log_{10}(T_e)$ of the form
\begin{equation}
S(x,y) = \sum_{i=1}^{N_x}\sum_{j=1}^{N_y} c_{ij}\, B_i(x)\, B_j(y),
\end{equation}
where $B_i$ and $B_j$ denote one-dimensional B-spline basis functions in $\log_{10}(n_e)$ and $\log_{10}(T_e)$, respectively, and $\{c_{ij}\}$ are the fitted spline coefficients. In the present work, we employ cubic B-splines with $n_{\mathrm{break}}=9$ breakpoints in $\log_{10}(n_e)$ and $n_{\mathrm{break}}=50$ breakpoints in $\log_{10}(T_e)$. These breakpoint counts are selected empirically to balance fitting accuracy and coefficient table size.

Along the temperature axis, breakpoint locations are chosen using adaptive strategies that concentrate knots in regions where the target plasma quantity varies rapidly with $T_e$. The candidate strategies include uniform spacing and several data-driven approaches based on estimated gradients, curvature, or residual indicators. In the first pass, spline fits are constructed using each candidate breakpoint strategy, and the strategy that yields the highest $R^2$ score against the tabulated data is selected for further refinement. To further improve accuracy, a two-pass refinement procedure is then applied using this selected strategy. An initial fit is constructed, and the fit is evaluated at all tabulated $(n_e,T_e)$ points, with the mean relative error computed for each temperature column. Additional breakpoints are inserted at temperature locations with the largest residuals, subject to a minimum separation from existing breakpoints. In the present implementation, five such additional temperature breakpoints are introduced during this refinement step, resulting in a total of 55 temperature breakpoints in the final spline model. The fit is then repeated using the enriched breakpoint set. This residual-driven enrichment improves the representation of localized structure in $T_e$ while maintaining a compact global model.

For radiative power loss, the spline is constructed for $\log_{10}(p_{\text{rad}})$, and the physical value is recovered by exponentiation. For $Z_{\text{avg}}$ and $Z_{\text{eff}}$, the spline is fitted in linear space. The spline coefficients are obtained from a weighted linear least-squares fit. The weighting is chosen to maintain accuracy in low-magnitude regions and to emphasize temperature intervals where strong gradients are present, thereby improving agreement in regimes that are most sensitive for coupled modeling.

After determining the spline coefficients, the quality of the surrogate representation is assessed using two standard validation metrics: the mean squared error (MSE) and the $R^2$ score. The MSE is defined as
\begin{equation}
\text{MSE} = \frac{1}{n} \sum_{i=1}^{n} (y_i - y'_i)^2,
\end{equation}
where $y_i$ and $y'_i$ denote the tabulated and predicted values, respectively, and $n$ is the number of data points. The $R^2$ score is defined as
\begin{equation}
R^2 = 1 - \frac{\sum_{i=1}^{n} (y_i - y'_i)^2}{\sum_{i=1}^{n} (y_i - \bar{y})^2},
\end{equation}
where $\bar{y}$ is the mean of the tabulated values. The $R^2$ score ranges from 0 to 1, with larger values indicating better agreement between the spline representation and the tabulated data. These metrics are computed over the full dataset used to determine the spline coefficients, and the results are summarized in Table~\ref{Table:cubic}. For radiative power loss, both the fitting and the reported metrics are evaluated in $\log_{10}(p_{\text{rad}})$ space to ensure a consistent dynamic range across parameters.

\begin{table}[!h]
\caption{\label{Table:cubic} Comparison of mean squared error (MSE) and $R^2$ scores for the B-spline surface fits of the CR plasma parameters for hydrogen, helium, neon, and argon, evaluated on the complete dataset. For the radiative power loss, the fit and the reported metrics are computed using $\log_{10}(p_{\text{rad}})$.}
\centering
\setlength{\tabcolsep}{10pt}
\begin{tabular}{c|cc|cc|cc}
\hline
\multirow{2}{*}{Species} 
& \multicolumn{2}{c|}{$p_{\text{rad}}$} 
& \multicolumn{2}{c|}{$Z_{\text{avg}}$} 
& \multicolumn{2}{c}{$Z_{\text{eff}}$} \\
\cline{2-7}
& MSE & $R^2$ & MSE & $R^2$ & MSE & $R^2$ \\
\hline
H  & 1.47882e-09 & 1.000000
   & 3.26089e-06  & 0.999481
   & -- & -- \\
He & 4.98648e-07 & 0.999996 
   & 1.26629e-08 & 0.999999 
   & 5.84628e-09 & 1.000000 \\
Ne & 6.81582e-07 & 0.999999 
   & 7.29374e-06 & 0.999998
   & 2.41755e-06 & 0.999999 \\
Ar & 5.04476e-07 & 0.999999 
   & 7.61018e-06 & 1.000000 
   & 5.54751e-06 & 1.000000 \\
\hline
\end{tabular}
\end{table}

\begin{figure}[!h]
\centering
\includegraphics[scale=0.21]{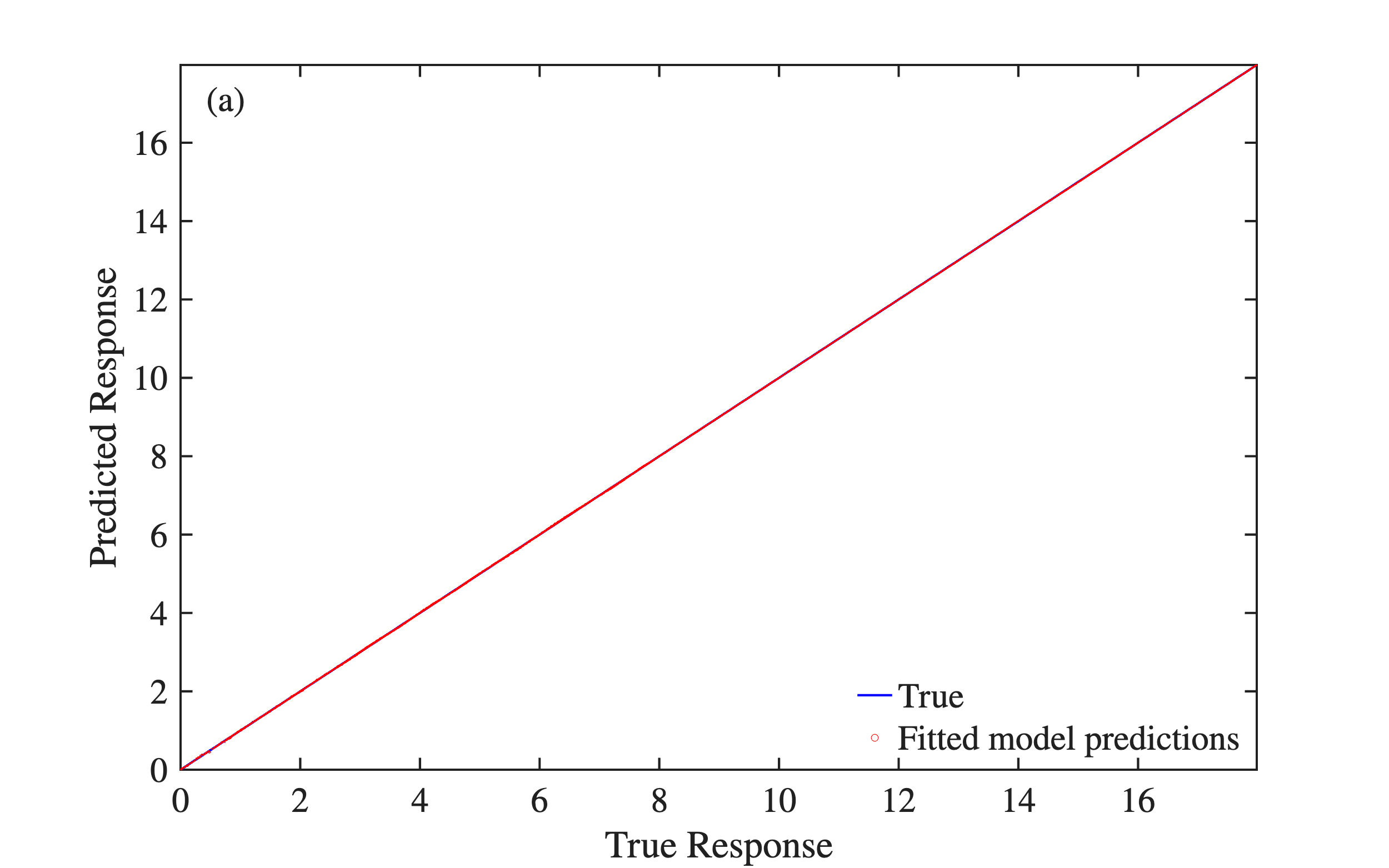}\includegraphics[scale=0.21]{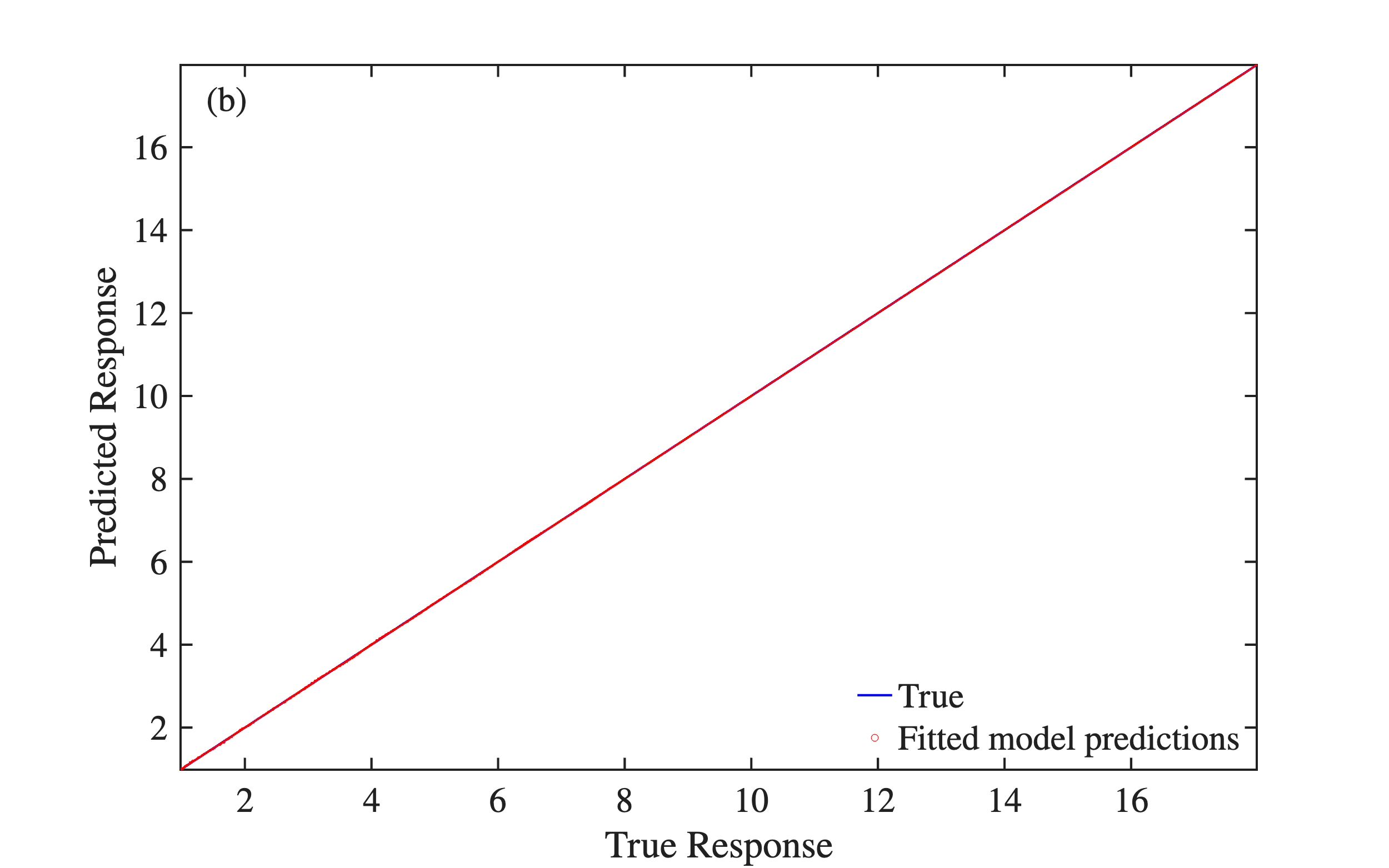} 
\includegraphics[scale=0.21]{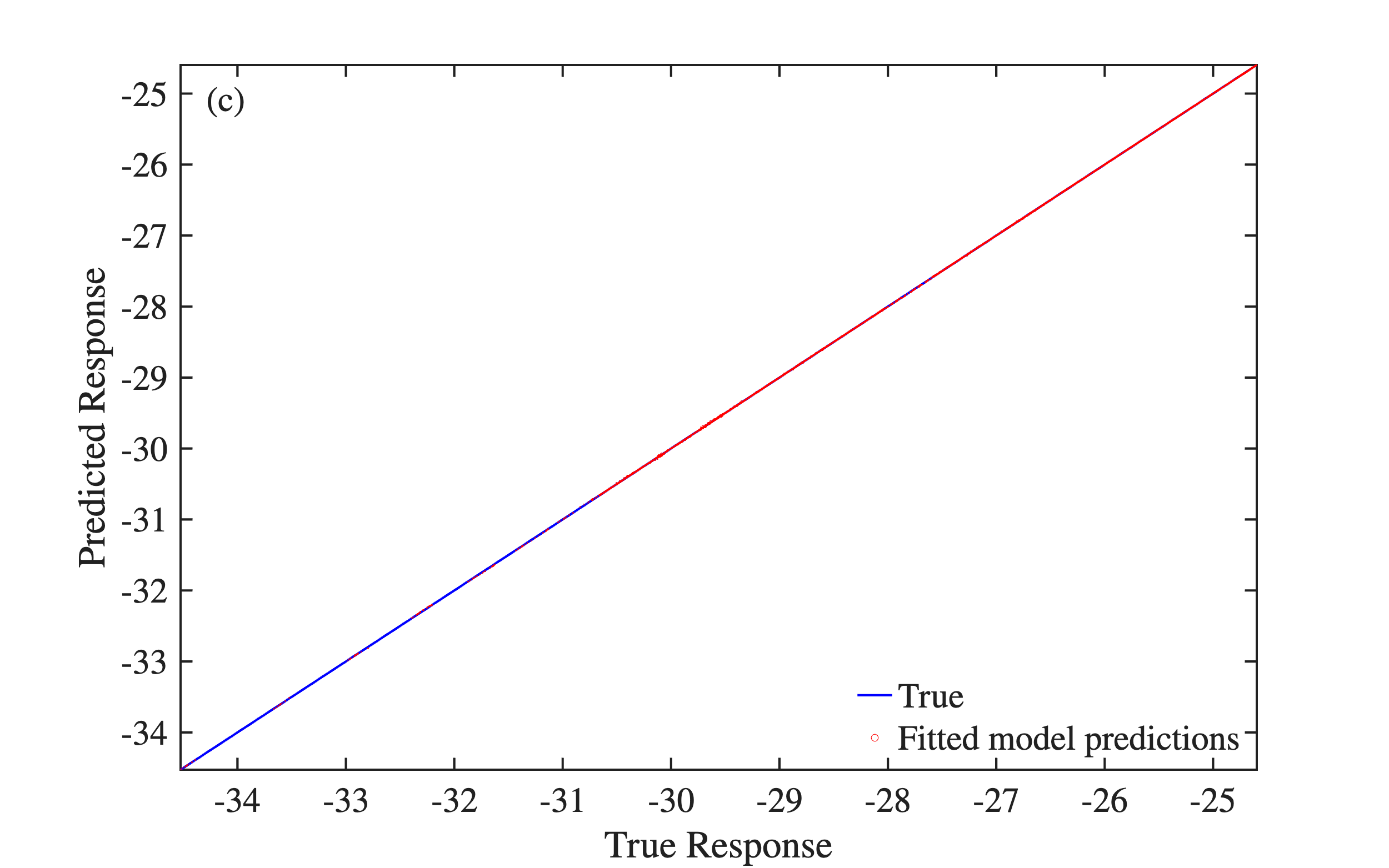} 
\caption{\label{Spline_Compare} Comparison of predicted and reference values obtained using the tensor-product B-spline surface representation: (a) average charge state, (b) effective charge state, and (c) $\log_{10}$ of the radiative power loss rate.}
\end{figure}

The results in Table~\ref{Table:cubic} indicate that the tensor-product spline representation reproduces the tabulated CR data with high accuracy across all species and plasma parameters considered. This approach is relatively simpler to incorporate into plasma simulation codes, as for each plasma species the data are contained within a single file of spline coefficients (along with the corresponding knot sequences and spline degree information). These files can be employed to extract plasma parameters (i.e., $p_{\text{rad}}$, $Z_{\text{avg}}$, and $Z_{\text{eff}}$) using three inputs: the plasma species identifier (used to select the corresponding spline-coefficient file), $T_e$, and $n_e$. The fidelity of the spline-based representation is further illustrated in Fig.~\ref{Spline_Compare} by comparing predicted values with the reference tabulated data over the full dataset used in this work. The strong correlation confirms that the fitted spline surfaces provide an accurate and smooth surrogate representation suitable for coupling to disruption-mitigation plasma models. 

\section{Conclusion}
In this work, we have presented a detailed analysis of the ionization balance and radiative power loss for plasmas relevant to tokamak disruption mitigation during the current quench phase. The calculations are carried out for four plasma species, i.e.,  hydrogen, helium, neon, and argon,  for tokamak relevant electron temperatures and electron densities, using the steady-state CR model, ATOMIC code, and recently developed FCR code. These results are compared with the predictions of the superconfiguration CE and CR models, and the common limitations of these simplified models are highlighted. Furthermore, we outline a simple yet effective method for coupling the CR data to plasma modeling, which allows the use of precomputed plasma parameters via a tensor-product B-spline surface representation. This approach enables the deployment of a high-fidelity CR model while maintaining computational efficiency. The results and comparisons presented in this paper provide valuable insights into the accuracy and applicability of different plasma models for tokamak disruption mitigation design and help guide future research in this critical area of fusion energy development.

\section{Acknowledgement}
This work was jointly supported by the US Department of Energy through the Fusion Theory Program of the Office of Fusion Energy Sciences and the SciDAC partnership on Tokamak Disruption Simulation between the Office of Fusion Energy Sciences and the Office of Advanced Scientific Computing, at Los Alamos National Laboratory (LANL) under Contract No. 89233218CNA000001.  C.J.F, M.C.Z, and J.C would like to specifically acknowledge Los Alamos National Laboratory (LANL) ASC PEM Atomic Physics Project.  LANL is operated by Triad National Security, LLC, for the National Nuclear Security Administration of the US Department of Energy (Contract No. 89233218CNA000001).  D.V.F. and I.B. acknowledge the support of the Australian Research Council, The National Computer Infrastructure, and the Pawsey Supercomputer Centre of Western Australia. This research used resources of the National Energy Research Scientific Computing Center, a DOE Office of Science User Facility supported by the Office of Science of the U.S. Department of Energy under Contract No. DE-AC02-05CH11231 using NERSC award FES-ERCAP0032298.

\section{Data Availability}
The C++ code and associated spline coefficient files are available from the authors upon reasonable request.

\bibliographystyle{apsrev}
\bibliography{reference.bib}

\end{document}